\documentclass{sig-alternate}

\usepackage[T1]{fontenc}
\usepackage[scaled=0.81]{beramono} 

\usepackage{amsfonts}
\usepackage{amsmath}
\usepackage{calc}
\usepackage{caption}
\usepackage{etoolbox} % for \undef
\usepackage{fancybox}
\usepackage{float}
\usepackage{flushend}
\usepackage{listings}
\usepackage[spacing=true]{microtype}
\usepackage{oosc2eiffel}
\usepackage{semantic}
\usepackage{subcaption}
\usepackage{tikz}
\usepackage{url}
\definecolor{darkblue}{rgb}{0,0,0.4}
\usepackage[colorlinks=true,allcolors=darkblue,breaklinks,draft=false,draft]{hyperref} 
\usetikzlibrary{calc}
\usetikzlibrary{shapes}
\usetikzlibrary{shapes.callouts}
\usetikzlibrary{backgrounds}
\usetikzlibrary{automata, arrows, chains, shadows}
\newtoggle{LONGVERSION}
\toggletrue{LONGVERSION}

\def\fig#1{Fig.\,}

\makeatletter
\newenvironment{centerbox}
   {\begin{Sbox}}
   {\end{Sbox}\centerline{\parbox{\wd\@Sbox}{\TheSbox}}}
\makeatother

\newcommand{\qsname}{SCOOP/Qs}
\newcommand{\rname}[1]{\ensuremath{\mbox{\textsc{#1}}}}
\newcommand{\bname}[1]{\textsf{#1}}
\newcommand{\smallsim}{{\raise.17ex\hbox{$\scriptstyle\sim$}}}

\newcommand{\lstinlinelang}[2]
  {\lstinline[language=#1, breaklines=false,%
    basicstyle=\ttfamily\normalsize]%
    {#2}%
  }

\newcommand{\eif}[1]{\lstinlinelang{oosc2eiffel}{#1}}
\newcommand{\lstinlinec}[1]{\lstinlinelang{C}{#1}}

\newcommand{\figref}[1]{\fig~\ref{fig:#1}}
\newcommand{\figlabel}[1]{\label{fig:#1}}

\newcommand{\sepr}{\eif{separate}}

\begin{document}

\toappear{}

\setlength{\paperwidth}{8.5in}
\setlength{\paperheight}{11in}
\setlength{\pdfpageheight}{\paperheight}
\setlength{\pdfpagewidth}{\paperwidth}

\title{Efficient and Reasonable Object-Oriented Concurrency}
\numberofauthors{2}
\author{
\alignauthor Scott West\titlenote{All research was done while employed at ETH Z\"{u}rich; opinions in this paper do not necessarily reflect those of Google Inc.}\\
\affaddr{Google Inc., Switzerland}\\
\email{scottgw@google.com}\\
\alignauthor Sebastian Nanz \qquad Bertrand Meyer\\
\affaddr{Department of Computer Science}\\
\affaddr{ETH Z\"{u}rich, Switzerland}\\
\email{firstname.lastname@inf.ethz.ch}
}

\maketitle

\begin{abstract}
  Making threaded programs safe and easy to reason about is one of the
  chief difficulties in modern programming.
  This work provides an efficient execution model for SCOOP,
  a concurrency approach that provides not only data-race freedom but also
  pre/postcondition reasoning guarantees between threads.
  The extensions we propose influence both the underlying semantics
  to increase the amount of concurrent execution that is possible,
  exclude certain classes of deadlocks,
  and enable greater performance.
  These extensions are used as the basis of
  an efficient runtime and optimization pass that
  improve performance 15$\times$ over a baseline implementation.
  This new implementation of SCOOP is, on average, also
  2$\times$ faster than other well-known safe concurrent languages.
  % Additionally, compilation and runtime strategies are developed
  % to efficiently execute SCOOP programs to the point where
  % the performance sees a 15$\times$ average improvement over baseline,
  % and 2$\times$ faster on average than other established concurrent languages
  % that also offer data race freedom.
  The measurements are based on both coordination-intensive
  and data-manipulation-intensive benchmarks designed
  to offer a mixture of workloads.
\end{abstract}



%{{{ introduction
\section{Introduction}

Programming languages and libraries that help programmers write
concurrent programs are the subject of intensive
research. Increasingly, special attention is paid to developing
approaches that provide certain execution guarantees; they support the
programmer in avoiding delicate concurrency errors such as data races
or deadlocks. For example, languages such as
Erlang~\cite{Armstrong:1996:CPE:229883} and others based on the Actor
model~\cite{Hewitt:1973:UMA:1624775.1624804} avoid data races by a
pure message-passing approach; languages such as
Haskell~\cite{PeytonJones:1996:CH:237721.237794} are based on Software
Transactional Memory~\cite{Shavit:1995:STM:224964.224987}, avoiding
some of the pitfalls associated with traditional locks.

Providing these guarantees can, however, be at odds with attaining good
performance. 
Pure message-passing approaches face the difficulty of how
to transfer data efficiently between actors;
and optimistic approaches to shared memory access,
such as transactional memory,
have to deal with recording, committing, and rolling back changes
to memory.
For this reason, execution strategies have to be developed
that preserve the performance of the language while maintaining
the strong execution guarantees of the model.

This work focuses on SCOOP~\cite{nienaltowski:2007:SCOOP},
an object-oriented approach to concurrency that aims to make concurrent
programming simpler by providing higher-level primitives that are more
amenable to standard programming techniques, such as pre/postcondition
reasoning. To achieve this goal, SCOOP places restrictions on the way
concurrent programs execute, thereby gaining more reasoning
capabilities but also introducing performance bottlenecks. To improve
the performance of SCOOP programs while maintaining the core of the
execution guarantees, this paper introduces a new execution model
called {\qsname}\footnote{Qs is pronounced ``queues'', as queues feature
  prominently in our new approach; the runtime and compiler
  associated with Qs is called Quicksilver,
  available from~\cite{west:2014:github_quicksilver}.}.
We first give
a formulation of the SCOOP semantics
which admits more concurrent behaviour than the existing
formalizations~\cite{morandi-et-al:2013:prototyping},
while still providing
the reasoning guarantees.
On this basis,
lower-level implementation techniques are developed to make the
scheduling and interactions between threads efficient.
These techniques are applied in %both in 
an advanced prototype implementation~\cite{west:2014:github_quicksilver}. %,
% as well as incorporated into
% the research branch of the
% production EiffelStudio compiler~\cite{se:2014:eve},
% the reference compiler for SCOOP programs.

The design and implementation choices are
evaluated on a benchmark suite that includes
computation-intensive and coordination-intensive workloads, showing
the advantages of the {\qsname} execution strategies. The
overall performance is compared to a broad variety of other paradigms
for parallel and concurrent programming --~C++/TBB, Go, Haskell, and
Erlang~-- demonstrating SCOOP's competitiveness.

The remainder of this paper is structured as follows.
Section~\ref{sec:model} introduces SCOOP and formally
specifies executions.
Section~\ref{sec:implementation} describes the implementation
techniques for this model.
Section~\ref{sec:optimization_evaluation} evaluates the effectiveness
of the different optimizations.
Section~\ref{sec:language_comparison}
compares {\qsname} against a variety of other paradigms. 
An analysis of related work is performed in
Section~\ref{sec:related-work}, and conclusions are drawn in
Section~\ref{sec:conclusion}.
%}}}

%{{{ Model

\section{Execution model}\label{sec:model}
The key motivation behind SCOOP~\cite{nienaltowski:2007:SCOOP} is
providing a concurrent programming model that allows the same kinds of
reasoning techniques that sequential models enjoy.  In particular,
SCOOP aims to provide areas of code where pre/postcondition reasoning
exists between independent threads.  To do this, SCOOP allows one to
mark sections of code where, although threads are operating
concurrently, data races are excluded entirely.

\subsection{A Brief Overview}

In \fig~\ref{fig:simple_scoop} one can see two programs that
are running in parallel.
\begin{figure}[htb]\centering
\small
\begin{minipage}{0.45\linewidth}
  \begin{lstlisting}
separate x
  do
    x.foo()
    a := long_comp()
    x.bar()
  end
  \end{lstlisting}
\centering
Thread 1
\end{minipage}
\begin{minipage}{0.45\linewidth}
  \begin{lstlisting}
separate x
  do
    x.bar()
    b := short_comp()
    c := x.baz()
  end
  \end{lstlisting}
\centering Thread 2
\end{minipage}  

\caption{A simple SCOOP program}
\label{fig:simple_scoop}
\end{figure}
Supposing that \eif{x} is the same object in each thread,
there are only two possible interleavings:
\begin{itemize}
\setlength{\itemsep}{0.3ex}
\item \eif{x.foo(), x.bar(), x.bar(), x.baz()} or
\item \eif{x.bar(), x.baz(), x.foo(), x.bar()}
\end{itemize}
However, in contrast to \eif{synchronized} blocks in Java,
these  \eif{separate} blocks not only protect access to shared memory,
but also initiate concurrent actions:
for both threads, the calls on \eif{x} are performed asynchronously,
thus for Thread 1, \mbox{\eif{x.foo()}} can execute in parallel
with \eif{long_comp()}.
However, it \emph{cannot} be executed in parallel with \eif{x.bar()}
as they have the same target, \eif{x}.
SCOOP has another basic operation, the query,
that provides synchronous calls.
It is so called because the sender expects an answer from
the other thread;
this is the case with the \eif{c := x.baz()} operation,
where Thread~2 waits for \eif{x.baz()} to complete before storing the
result in \eif{c}.

The SCOOP model associates every object with a thread of execution,
its \emph{handler}.
There can be many objects associated to a single handler,
but every object has exactly one handler.
In \fig~\ref{fig:simple_scoop}, \eif{x} has a handler that
takes requests from Threads~1 and~2.
The threads that wish to send requests to \eif{x} must register
this desire,
which is expressed in the code by \mbox{\eif{separate x}.}
The threads are deregistered at the end of the \eif{separate} block.

This model is similar to other message passing models, such as the
Actor model~\cite{Hewitt:1973:UMA:1624775.1624804}.  What
distinguishes SCOOP from languages like
Erlang~\cite{Armstrong:1996:CPE:229883} is that the threads have more
control over the order in which the receiver will process the messages.
When multiple processes each send multiple messages to a single
receiver in Erlang, the sending processes do not know the order of
processing of their messages (as they may be interleaved with messages from
other processes).
In SCOOP, since each thread registers with the receiver,
the messages from a single separate block to its handler will be processed in order,
without any interleaving.

This ordering gives the programmer the ability to reason about
concurrent programs in a sequential way within the \eif{separate} blocks.
To be precise, pre/postcondition reasoning can be applied to
a \eif{separate} object protected by a \eif{separate} block,
even though the actions are being executed in parallel.
A \eif{separate} object is marked as such by the type system,
and methods may only be called on a \eif{separate} object if it
is protected by a \eif{separate} block.
Maintaining reasoning among multiple independent
\eif{separate} objects is also possible,
but requires all \eif{separate} objects concerned
be protected by the same \eif{separate} block.

\tikzstyle{queue_block} =
  [ draw=green!50!black!40
  , left color=green!50!black!40
  , rectangle
  , rounded corners=2pt
  , minimum size = 1em
  ]
\def\queueAt#1#2#3#4{
  \foreach \x in {0, 1,...,#4} {
    \node[queue_block] at (#1 em + \x em, #2 em) (q_\x_#3) {};
  }
}
The original SCOOP operational semantics~\cite{nienaltowski:2007:SCOOP} mandated
the use of a lock to ensure that pre/postcondition reasoning
could be applied by a client on its calls to a handler.
One can visualize this as the client $c_0$ placing the calls
in a queue for the handler $h$ to dequeue and process,
as in \fig~\ref{fig:normal_handler}.
\begin{figure}
\begin{center}
  \begin{tikzpicture}[rectangle, rounded corners=2pt,
     redbox/.style={draw=red!50, left color=red!80!black!20},
     bluebox/.style={draw=blue!50, left color=blue!50!black!20},
     greenbox/.style={draw=green!50!black!40, left color=green!50!black!40}]
    \queueAt{0}{1}{0}{4}

    \node[draw, dashed, right=0em of q_4_0, minimum size=1em] (outgoing) {};
    \node[draw, redbox, above=1em of q_2_0] (h) {h};
    \node[draw, greenbox, minimum size=1em, right=0.2em of h] (old) {};
    \path[draw, -stealth, dashed]
        (outgoing.north) to[out=90, in=0] (old.east);

    % The present client, the work its enqueueing, and where the work is going
    \node[draw, dashed, left=0em of q_0_0, minimum size=1em] (incoming) {};
    \node[bluebox, below=1em of q_2_0] (c0) {$c_0$};
    \node[greenbox, minimum size=1em, left=0.2em of c0] (new) {};
    \path[draw, -stealth, dashed] (new.west) to[out=180, in=270] (incoming.south);

    % The lock
    \path[draw=brown,
          thick,
          left color=orange!90!black!50,
          right color=orange!50!black!10]
                  ($(incoming.south west) + (-1em,-0.5em)$) --
                ++(-1.5em, 0) --
                ++(0, 1.5em) node (upleft) {} --
                ++(1.5em, 0) node (upright) {} --
                ++(0, -1.5em) --
                cycle;
   \draw[line width=0.2em]
       ($(upleft.center) + (0.2em,0em)$) ..
         controls ($(upleft.center) + (0.2em, 1em)$) and
                  ($(upright.center) + (-0.2em, 1em)$)
    .. ($(upright.center) + (-0.2em, 0em)$);

    % The waiting clients
    \node[bluebox, left=4em of q_0_0] (c1) {$c_1$};
    \node[bluebox, left=0.5em of c1] (c2) {$c_2$};
    \node[bluebox, left=0.5em of c2] (c3) {$c_3$};
  \end{tikzpicture}
\end{center}
\caption{Normal handler implementation}
\label{fig:normal_handler}
\end{figure}
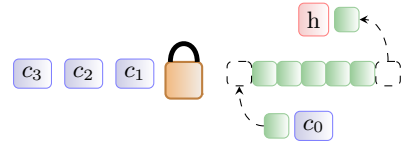
The other clients ($c_1, c_2, c_3$) that may want to access the handler's queue
must wait until the current client is finished.

\subsection{Reasoning Guarantees}

There are a few key reasoning guarantees that
an implementation of SCOOP must provide:

\begin{enumerate}
\setlength{\itemsep}{0.3ex}
\item Regular (non-\eif{separate}) calls
  and primitive instructions (assignment, etc.)
  execute immediately and are synchronous.
\item
  \begin{sloppypar}
  Calls to another handler, \eif{h}, on which object \eif{x} resides,
  within the body of a \eif{separate x} block
  will be executed in the order they are logged,
  and there will be no intervening calls logged
  from other clients.
\end{sloppypar}
\end{enumerate}

The effect of rule 1 is that normal sequential reasoning is applied
to calls that are issued by the client, to the client.
Rule 2 implies that calls that are made from the client to the handler
are applied in order,
thus the client can apply pre-/postcondition
reasoning from one call it has made to the next.

\def\this{\mathtt{this}}
\def\send{\mathtt{call}}
\def\query{\mathtt{query}}
\def\access{\mathtt{access}}
\def\order{\mathit{order}}
\def\ordernorm{\order_{P, H,\this}}
\def\events{\mathit{events}}
\def\procorder{\overrightarrow{\sqsubset}}
\def\handler{\mathtt{handler}}

\def\handler#1#2#3{\left(#1, #2, #3\right)}
\def\onelinerule#1#2{#1 => #2}
\def\splitrule#1#2{
  \arraycolsep=1.4pt\def\arraystretch{1.1}
  \begin{array}{l}
    #1 => \\
    #2
  \end{array}
  \undef\arraystretch
}
\def\parsplitrule#1#2#3#4{
  \arraycolsep=1.4pt\def\arraystretch{1.1}
  \begin{array}{lcl}
    #1&||&#2 => \\
    #3&||&#4 
  \end{array}
  \undef\arraystretch
}

\begin{figure*}[!htbp]
\small
\[
\begin{array}{c}
\inference[\rname{separate}]
{}
{\parsplitrule
  {\handler{h}{q_h}{\mathtt{separate}\ x\ s}}
  {\handler{x}{q_x}{t}}
  {\handler{h}{q_h}{s; \send(x, \mathsf{end})}}
  {\handler{x}{q_x + \left[ h \mapsto [] \right]}{t}}
} \qquad

\inference[\rname{call}]
{}
{\parsplitrule
  {\handler{h}{q_h}{\send(x, f)}}
  {\handler{x}{q_x}{t}}
  {\handler{h}{q_h}{\mathsf{skip}}}
  {\handler{x}{q_x\left[h \mapsto q_x[h] + [f] \right]}{t}}
} \\\\

\inference[\rname{query}]
{}
{\parsplitrule
  {\handler{h}{q_h}{\query(x, f)}}
  {\handler{x}{q_x}{t}}
  {\handler{h}{q_h}{\mathsf{wait}\ x}}
  {\handler{x}{q_x\left[h \mapsto q_x[h] + 
        \left[f, \mathsf{release}\ h\right] \right]}{t}}
} \qquad

\inference[\rname{sync}]
{}
{\parsplitrule
  {\handler{h}{q_h}{\mathsf{wait}\ x}}
  {\handler{x}{q_x}{\mathsf{release}\ h}}
  {\handler{h}{q_h}{\mathsf{skip}}}
  {\handler{x}{q_x}{\mathsf{skip}}}
} \\\\

\inference[\rname{run}]
{}
{\splitrule
  {\handler{h}
    {\left[x \mapsto \left[s\right] + \mathit{ss}\right] +
      \mathit{ys}}{\mathsf{skip}}}
  {\handler{h}{\left[x \mapsto \mathit{ss}\right] + \mathit{ys}}{s}}
} \qquad

\inference[\rname{end}]
{}
{\splitrule
  {\handler{h}{\left[x \mapsto []\right] + \mathit{ys}}{\mathtt{end}}}
  {\handler{h}{\mathit{ys}}{\mathsf{skip}}}
} \qquad

\inference[\rname{seq}]
{\onelinerule
  {\handler{h}{\mathit{xs}}{\mathit{s_1}}}
  {\handler{h}{\mathit{xs}}{\mathit{s_1'}}}
}
{\onelinerule
  {\handler{h}{\mathit{xs}}{\mathit{s_1 ; s_2}}}
  {\handler{h}{\mathit{xs}}{\mathit{s_1' ; s_2}}}
} \\\\

\inference[\rname{seqSkip}]
{}
{\onelinerule
  {\handler{h}{xs}{\mathsf{skip} ; s_2}}
  {\handler{h}{xs}{\mathit{s_2}}}
} \qquad

% \inference[\rname{parStep}$_1$]
% {\onelinerule{P}{P'}}
% {\onelinerule
%   {P\ ||\ Q}
%   {P'\ ||\ Q}
% } \qquad

% \inference[\rname{parStep}$_2$]
% {\onelinerule{Q}{Q'}}
% {\onelinerule
%   {P\ ||\ Q}
%   {P\ ||\ Q'}
% } \\\\

\inference[\rname{parStep}]
{\onelinerule{Q}{Q'}}
{\onelinerule
  {P\ ||\ Q}
  {P\ ||\ Q'}
} \\\\

\inference[\rname{oneStep}]
{\onelinerule{P}{Q}}
 {P =>^{*} Q}
\qquad
\inference[\rname{manyStep}]
{P =>^{*} P' & P' =>^{*} Q}
{P =>^{*} Q}
\end{array}
\]
\caption{Inference rules of {\qsname}}
\label{fig:new_operational_rules}
\end{figure*}

\subsection{The {\qsname} Execution Model}
\label{sec:op-sem}

The first SCOOP guarantee is easy to achieve,
it is simply how sequential programs operate.
To understand how to implement SCOOP efficiently,
it is important to concentrate on the the second guarantee.
This guarantee states that the requests from a particular
client are processed by the handler in the order they are sent,
disallowing interleaving requests from other clients.
To prevent clients from interfering with one another
on a particular handler can be
achieved by giving each client their own private area (a queue)
in which to place their requests.
Each client then just shares this private queue with the handler
to which it wants to send requests.

\paragraph{Syntax} The following syntax of statements $s$ is used to describe the execution model.
\begin{displaymath}
  \begin{array}{rl}
    s ::= & \mathtt{separate}\ x\ s\ |\ \send(x, f)\ |\ \query(x, f)\ |\ \\
          & \mathsf{wait}\ h\ |\ \mathsf{release}\ h\ |\ \mathsf{end}\ |\ \mathsf{skip}
  \end{array}
\end{displaymath}
Note that $\mathtt{separate}$ blocks and $\send$ and
$\query$ requests model instructions of SCOOP programs,
whereas the statements $\mathsf{wait}$, $\mathsf{release}$,
$\mathsf{end}$, and $\mathsf{skip}$ are only used to model the runtime
behaviour. In particular, statements $\mathsf{wait}$ and
$\mathsf{release}$ describe the synchronization to wait for the result
after a $\mathtt{query}$, statement $\mathsf{end}$ models the end of
a group of requests, and $\mathsf{skip}$ models no behaviour.

\paragraph{Operational Semantics} In
\fig~\ref{fig:new_operational_rules}, an operational semantics that
conforms to the SCOOP guarantees is given.
It is described using inference rules for transitions of the form $P => Q$,
where $P$ and $Q$ are parallel
compositions of handlers.
The $||$ operator is commutative and associative to facilitate
appropriate reordering of handlers.

The basic representation of a handler is a triple $(h, q_h, s)$ of its
identity $h$, request queue $q_h$, and the current program it is executing, $s$. 
A request queue is a list of handler-tagged private queues,
and is thus really a queue-of-queues.
Private queues of a client handler $c$ can be looked-up $q_h[c]$,
and can be updated $q_h[c \mapsto l]$,
where $l$ is the new list to associate with the handler $h$.
Both lookup and updating work on the \emph{last} occurrence of $c$,
which is important as this is the one that the client modifies.
The queue can also be decomposed structurally,
with $[x \mapsto s] + ys$ meaning that the head of the queue
is from client $x$ with private queue $s$,
and $ys$ is the rest of the structure (possibly empty).
So although the private queues in the queue-of-queues can be
accessed and modified in any order,
they are inserted and removed in first-in-first-out order.

We describe the unique operations of
\fig~\ref{fig:new_operational_rules}: \mbox{\eif{separate}} blocks (the
rule \rname{separate}), the two different kinds of requests
(\rname{call}, \rname{query}, \rname{sync} rules), and how these
requests are processed by the handlers (\rname{run} and \rname{end}
rules). The sequential and parallel composition rules are defined in
the standard way.

In the rule \rname{separate},
clients insert their private queue at the end of the handler's request queue.
This operation occurs at the beginning of a \eif{separate} block.
This registers them with the handler,
who will eventually process
the requests.
The fact that a handler only processes one private queue at a time
ensures that the reasoning guarantees are maintained. % replaces the previous locking behaviour.
It is also a completely asynchronous operation,
as the supplier's handler-triple only consists of variables, 
i.e., there are no restrictions on what state the supplier has to be in
for this rule to apply.
Additionally, the client appends a $\send(x, \texttt{end})$ action
before the end of the \eif{separate} block to signal that the
supplier $x$ can take requests from other clients.

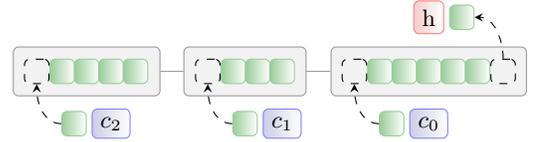
\begin{figure}[htb]
\begin{center}
  \begin{tikzpicture}[rounded corners=2pt,
    redbox/.style={draw=red!50, left color=red!80!black!20},
    bluebox/.style={draw=blue!50, left color=blue!50!black!20},
    greenbox/.style={draw=green!50!black!40, left color=green!50!black!40}]

    \queueAt{0}{1}{0}{4}

    % The present client, the work its enqueueing, and where the work is going
    \node[draw, dashed, left=0em of q_0_0, minimum size=1em] (incoming0) {};
    \node[bluebox, below=1em of q_2_0] (c0) {$c_0$};
    \node[greenbox, minimum size=1em, left=0.2em of c0] (new) {};
    \path[draw, -stealth, dashed]
       (new.west) to[out=180, in=270] (incoming0.south);

    \node[draw, dashed, right=0em of q_4_0, minimum size=1em] (outgoing) {};
    \node[draw, redbox, above=1em of q_2_0] (h) {h};
    \node[draw, greenbox, minimum size=1em, right=0.2em of h] (old) {};
    \path[draw, -stealth, dashed]
        (outgoing.north) to[out=90, in=0] (old.east);

    \begin{scope}[on background layer]
      \node[rectangle, minimum width=8em, minimum height=2em,
            draw=black!40, fill=black!5]
        (r0) at (q_2_0) {};
    \end{scope}

    % Client c1, the work its enqueueing, and where the work is going
    \queueAt{-6}{1}{0}{2}

    \node[draw, dashed, left=0em of q_0_0, minimum size=1em] (incoming) {};
    \node[bluebox, below=1em of q_2_0] (c) {$c_1$};
    \node[greenbox, minimum size=1em, left=0.2em of c] (new) {};
    \path[draw, -stealth, dashed]
      (new.west) to[out=180, in=270] (incoming.south);

    \begin{scope}[on background layer]
      \node[rectangle, minimum width=5em, minimum height=2em,
            draw=black!40, fill=black!5]
        (r1) at (q_1_0.west) {};
    \end{scope}

    % Client c1, the work its enqueueing, and where the work is going
    \queueAt{-13}{1}{0}{3}

    \node[draw, dashed, left=0em of q_0_0, minimum size=1em] (incoming) {};
    \node[bluebox, below=1em of q_2_0] (c) {$c_2$};
    \node[greenbox, minimum size=1em, left=0.2em of c] (new) {};
    \path[draw, -stealth, dashed]
      (new.west) to[out=180, in=270] (incoming.south);

    \begin{scope}[on background layer]
      \node[rectangle, minimum width=6em, minimum height=2em,
            draw=black!40, fill=black!5]
        (r2) at (q_1_0) {};
    \end{scope}

    % \node[above=1em of r2] (qoq_callout) {queue of queues};
    % \path[draw] (qoq_callout.south) -- (r2.north);

    % \node[above=1em of r1] (pq_callout) {private queue};
    % \path[draw] (pq_callout.south west) -- (q_3_0.north);

    \path[draw, draw=black!40] (r0.west) -- (r1.east);
    \path[draw, draw=black!40] (r1.west) -- (r2.east);
  \end{tikzpicture}
\end{center}
\caption{Handler implementation based on queue of queues}
\label{fig:qoq_handler}
\end{figure}

The {\qsname} semantics,
in contrast to the original lock-based SCOOP semantics,
uses multiple queues that can all be accessed and
enqueued into simultaneously by clients.
This behaviour is visualized in \fig~\ref{fig:qoq_handler},
where the outer (gray) boxes are nodes in the queue of queues,
and the inner (green) boxes are nodes in the private queues.
This nested queueing maintains the reasoning
guarantees while still
allowing all clients to enqueue
asynchronous calls without waiting.

In rule \rname{call}, the $\send$ action is non-blocking:
it asynchronously appends the requested method $f$ to the end of the 
appropriate client's private queue. %, raising a $\send_h(x, f)$ event.

Rule \rname{query}, requesting a query execution, however, does require blocking
as it must wait for the result of the function application.
This is modeled by sending the query request %, raising a $\query_h(x, f)$ event,
and introducing a pair of actions which can only step forward together:
the \textsf{wait}/\textsf{release} pair.
There is only one rule (\rname{sync}) that can rewrite these into \textsf{skip},
and it can only do so when both processes are executing each of the pair.
%, raising an $\access_h$ event.

Each handler processes its request queue in the following way:
in rule \rname{run}, if the handler is idle (executing \textsf{skip}) then
it will examine the request queue.
If the request queue's first entry (a private queue) is non-empty, then the first
action is taken out of that private queue and placed in the program part
of the handler to execute.
If the request queue is empty, or it contains an empty private queue
as its first entry, then the thread does nothing.
In rule \rname{end}, the thread finishes one private queue and switches to the next
when it encounters the \textsf{end} statement,
which was placed by the owner of the outgoing private queue
when it finished executing its \eif{separate} block (rule \rname{separate}).

%%%
%%% I would like to toggle this out but it doesn't interact well with the listing.
%%%
% \iftoggle{LONGVERSION}
% {
% \subsection{Deadlock reduction}
% The {\qsname} semantics presented above also provides a reduction
% in the number of deadlock situations that can occur.
% In particular, since the original semantics of
% handler acquisition was blocking,
% one could construct programs that would deadlock
% merely trying to acquire the appropriate handlers.
% For example, the program  in \fig~\ref{fig:old_deadlock}
% %
% \begin{figure}[htb]
%   \small
% \begin{minipage}{0.49\linewidth}
% \begin{lstlisting}
% separate x
%   separate y
%     x.foo()
%     y.bar()
%   end
% end
% \end{lstlisting}
% \centering Client 1
% \end{minipage}
% \begin{minipage}{0.49\linewidth}
% \begin{lstlisting}
% separate y
%   separate x
%     x.foo()
%     y.bar()
%   end
% end
% \end{lstlisting}
% \centering Client 2
% \end{minipage}
% \caption{Possible deadlock situation}
% \label{fig:old_deadlock}
% \end{figure}
% %
% will deadlock under some schedules when using
% the original semantics.
% This is due to the inconsistent locking order of \eif{x} and \eif{y}.
% However, in the {\qsname} execution model this example
% cannot deadlock because there are no longer any
% blocking operations:
% both clients can simultaneously reserve the handlers
% \eif{x} and \eif{y}, and log asynchronous calls on them.
% Preventing this manner of deadlock was
% the subject of previous research~\cite{west-et-al:2010:oo-deadlock}.
% }

\subsection{Multiple Handler Reservations}
The {\sepr} block as shown so far only reserves a single handler,
and this block provides race-freedom guarantees between
a single client and a single handler.
However, a client may want to ensure consistency
among multiple handlers or objects.
To provide guarantees about multiple handlers,
a multiple handler {\sepr} block must be used,
as in \figref{multi_reserve}.
\begin{figure}[htb]
  \centering
  \small
  \begin{subfigure}{0.45\linewidth}
    \begin{lstlisting}
separate x y
  do
    x.set (Red)
    y.set (Red)
  end    
    \end{lstlisting}
    \caption{Thread 1}
  \end{subfigure}
  \begin{subfigure}{0.45\linewidth}
    \begin{lstlisting}
separate x y
  do
    x.set (Blue)
    y.set (Blue)
  end
    \end{lstlisting}
    \caption{Thread 2}
  \end{subfigure}
  \caption{Multiple reservations}
  \figlabel{multi_reserve}
\end{figure}
In this example, this has the effect that, whenever a client
reserves both \eif{x} and \eif{y},
the colours of each object will be the same,
either both red or both blue.
When written in this way and executed under either SCOOP or SCOOP/Qs,
any client that comes after the execution of Thread~1 or Thread~2 (or both),
and reserves \eif{x} and \eif{y} together
will always see \eif{x.colour = y.colour}.
If using nested reservation,
this may not be the case due to a possible race enqueueing
the private queue into the queue-of-queues.

The modification to the \rname{separate} rule to support this 
is straight-forward.
First one defines an update function that updates a handler
if it is in the set $X$.
\[
\footnotesize
\begin{array}{l}
  %% If this isn't put in an array the footnotesize from above doesn't really work
\text{resOne} (X, h, \handler{x}{q_x}{t}) = \left\{
\begin{array}{lr}
   \handler{x}{q_x + \left[ h \mapsto [] \right]}{t} & \text{if}\ x \in X \\
   \handler{x}{q_x}{t} & \text{if}\ x \notin X \\
 \end{array}
\right.
\end{array}
\]
Then this is applied over the parallel composition of all handlers.
\[
\footnotesize
\begin{array}{lcl}
  \text{resMany}(X, h, P\ ||\ Q) & = & \text{resMany}(X, h, P)\ ||\\
                                 &   & \text{resMany}(X, h, Q) \\
  \text{resMany}(X, h, \handler{x}{q_x}{t}) & = & \text{resOne}(X, h, \handler{x}{q_x}{t})
 \end{array}
\]
Lastly, a function describes that each handler in the set
(represented here by a list so it can be traversed)
is sent an $\mathsf{end}$ message.
\[
\footnotesize
\begin{array}{lcl}
  \text{endMany}(x :: xs) & = & \send(x, \mathsf{end}) ; \text{endMany}(xs) \\
  \text{endMany}([]) & = & \textsf{skip} \\
 \end{array}
\]
These functions combine to define a generalized \rname{separate}
rule that can reserve multiple handlers atomically.
\[
\footnotesize
\inference[\rname{separate}]
{P' = \text{resMany}(X, h, P) \\
 \text{ends} = \text{endMany}(X)}
{\parsplitrule
  {\handler{h}{q_h}{\mathtt{separate}\ X\ s}}
  {P}
  {\handler{h}{q_h}{s; \text{ends}}}
  {P'}
}
\]

\subsection{Deadlock}
Under the original handler implementation of SCOOP,
the program  in \figref{old_deadlock}
\begin{figure}[htb]
\begin{minipage}{0.49\linewidth}
\begin{lstlisting}
separate x
  do
    separate y
      do
        x.foo()
        y.bar()
      end
  end
\end{lstlisting}
\centering Client 1
\end{minipage}
\begin{minipage}{0.49\linewidth}
\begin{lstlisting}
separate y
  do
    separate x
      do
        x.foo()
        y.bar()
      end
  end
\end{lstlisting}
\centering Client 2
\end{minipage}
\vspace{1ex}
\caption{Possible deadlock situation}
\label{fig:old_deadlock}
\end{figure}
will deadlock under some schedules.
This is due to the inconsistent locking
order of \eif{x} and \eif{y}.
However, in the {\qsname} execution model this example
cannot deadlock because there are no longer any
blocking operations:
both clients can simultaneously reserve the handlers
\eif{x} and \eif{y}, and log asynchronous calls on them.
Deadlock is still possible in {\qsname},
however one must also use queries (which block) to achieve
the same effect.
If \eif{x.query} and \eif{y.query} are added to the
innermost \eif{separate} blocks of Client 1 and Client 2, respectively,
the program may deadlock.

%}}}

%{{{ Implementation

\section{Implementation}\label{sec:implementation}

The semantics described in Section~\ref{sec:model} are used
to implement a compiler and runtime for SCOOP programs.
The operational semantics gives rise to
notable runtime performance and implementation properties.
We pay particular attention to how to move the implementation
from a synchronization-heavy model to one which
reduces the amount of blocking.
%  and is more suitable to facilitating
% distributing computations across a network,
% one of the original intentions for SCOOP.

The runtime for {\qsname} is written in C,
the compiler is written in Haskell and
targets the LLVM framework~\cite{llvm:2014:site} to take advantage of the
lower level optimizations that are available.
Using LLVM is a necessary choice for this work because
it is important to compare with other more mature languages and
the comparison should not focus on obvious shortcomings,
such as a lack of standard optimizations.
LLVM is also built to be extended;
this work extends LLVM by adding a custom optimization pass.
The {\qsname} compiler, runtime, and benchmarks are available from
GitHub~\cite{west:2014:github_quicksilver}.

The runtime is broken into 3 layers:
task switching,
light-weight threads,
and handlers.
Some of the optimizations described in this section
take place at the handler layer,
but there are also some that use
the other two layers as well
to optimize scheduling.

\subsection{Request Processing}

The $\rname{run}$ and $\rname{end}$ rules describe all of the
queue management facilities that a handler has to perform.
This correspondence is shown in the high-level implementation of the
main handler-loop given in \fig~\ref{fig:handler_loop}.

\begin{figure}[htb]
  \centering
  \footnotesize
  \begin{lstlisting}[language=C++, keepspaces=true]
// RUN rule, when there is a private queue
//   available
while (qoq.dequeue (&private_queue))
  {
    // if dequeue returns true:
    //    RUN rule; process calls from
    //    this queue.
    // otherwise:
    //    END rule; switch to the next
    //    private queue
    while (private_queue.dequeue (&call))
      {
         execute_call (call);
      }
  }
\end{lstlisting}
\caption{Main handler-loop}
\label{fig:handler_loop}
\end{figure}

The structure of the handler's loop directly corresponds to the
data structure implementation (a queue of queues).
One can see that private queues are continually taken from
the outer queue, where the dequeue operation returns a Boolean result.
False corresponds to no more work (indicating the processor can shut down),
not that the queue is empty as may be in a non-blocking
queue implementation.
For each private queue that is received,
calls are repeatedly dequeued out of it and executed
until false is returned from the dequeue operation,
indicating that the $\rname{end}$ rule has been triggered,
and the client presently does not wish to log more requests.

Note that the arrangement of
clients and handlers follows a particular pattern when
the queue-of-queues pattern is used.
Namely, that each handler first reserves a position
in the queue-of-queues:
each queue-of-queues has many clients trying to gain access,
but only one handler removing the private queues.
This is a typical multiple-producer single-consumer arrangement,
so an efficient lock-free queue specialized for this case 
can be used to implement the queue-of-queues.
Similarly, once the private queue has been dequeued by the
handler the communication is then single-producer single-consumer;
the client enqueues calls, the handler dequeues and executes them.
Again an efficient queue can be constructed to especially
handle this case.
These optimizations are important as they are
involved in all communication between
clients and handlers.

\subsection{Client Requests}
The handler-loop implementation, above,
resides in the runtime library.
The client-side is where the compilation and runtime system meet.
In particular, the compiler emits the code allowing the client to
package and enqueue requests for the handler,
and handle waiting for the results of \eif{separate} queries.

\begin{figure}[htb]
  \centering
  \footnotesize
  \begin{lstlisting}[language=C++]
private_queue* h_p = client.queue_for (h);

// SEPARATE rule, adding an empty queue
//   to the queue of queues
h.qoq.enqueue (h_p);

<compiled body>

// SEPARATE rule, compiler adds the
//   code to enqueue the END marker
h_p.enqueue (END);
\end{lstlisting}
  \caption{A compiled \lstinline{separate} block}
  \label{fig:compiled_separate}
\end{figure}
When a client reserves a handler with \lstinline{separate h do <body> end},
this corresponds to the code shown in \fig~\ref{fig:compiled_separate}.
The client receives a private queue \lstinline{h_p} for the desired handler $h$,
represented in the \rname{separate} rule by
the private queue appearing on the handler's queue-of-queues.
This private queue can either be freshly created or
taken from a cache of queues, to improve execution speed.
The client then enqueues this new private queue on
the queue-of-queues for the handler,
which means the private queue is now ready to log calls in the body.
Finally, corresponding to the end of the \lstinline{separate} block,
the constant denoting the end of requests is placed in the private queue,
allowing the handler to move on to the next client.

There will typically be calls to the handler in the body of a \lstinline{separate} block.
The asynchronous calls are packaged using
the libffi library~\cite{libffi:2014:site},
which abstracts away the details of various calling conventions.
This packaged call is then put into
the proper private queue for
the desired handler.
This can be seen in \fig~\ref{fig:compile_async_call},
\begin{figure}[htb]
  \centering
  \footnotesize
  \begin{lstlisting}[language=C++, keepspaces=true]
arg_types[0] = &ffi_type_pointer;
arg_values[0] = &arg;
ffi_prep_cif(ffi_call, FFI_DEFAULT_ABI, 1,
             &ffi_type_void, arg_types);

// CALL rule, showing the setup via libffi.
h_p.enqueue(call_new(ffi_call, 1, arg_values));
\end{lstlisting}
  \caption{Enqueueing an asynchronous call}
  \label{fig:compile_async_call}
\end{figure}
the enqueue operation relating directly to the
\rname{call} rule.
Packaging the call entails setting up the call interface (cif)
with the appropriate argument and return types with
\lstinline{ffi_prep_cif},
and then storing the actual arguments for later application by the  handler.
Note that the allocation of arguments and argument types for the call
cannot be done on the client's stack
because the call may be processed by the handler
after the client's stack frame has been popped.

For efficiency reasons,
a different strategy is used for synchronous calls (queries).
This is because packaging a call involves
allocating memory, populating structures,
and the handler must later unpack it.
In short: this takes longer than a regular function call.
In the asynchronous case these steps are unavoidable because
the execution of the call must be done in parallel with the client's operations.
However, for synchronous calls this is not the case:
the client will be waiting for a reply from the supplier
when the supplier finishes executing the query.
To make use of this optimization opportunity,
for shared-memory systems,
we can change the $\rname{query}$ rule to the following:

{
  \footnotesize
\[
\inference[]%\rname{queryNew}]
{}
{\parsplitrule
  {\handler{h}{q_h}{\query(x, f)}}
  {\handler{x}{q_x}{t}}
%  {\query_h(x,f)}
  {\handler{h}{q_h}{\mathsf{wait}\ x; f}}
  {\handler{x}{q_x\left[h \mapsto q_x[h] + 
        \left[\mathsf{release}\ h\right] \right]}{t}}
}
\]
}

Note that the execution of the call $f$ is shifted to the client,
after the synchronization with the handler has occurred.
This does not change the execution behaviour because,
as in the original rule,
all calls on the handler are processed before the query
and the client does not proceed to log more calls
until the query has finished executing.
As can be seen from \fig~\ref{fig:compile_query_call},
\begin{figure}[htb]
  \footnotesize
  \centering
\begin{subfigure}[b]{0.49\linewidth}
  \begin{lstlisting}[language=C++, keepspaces=true]
<packing same as async>
ffi_call(&ffi_call, f,
         &result, 0);
// QUERY rule
h_p.enqueue(ffi_call);
// SYNC rule
h_p.sync();
  \end{lstlisting}
\caption{\raggedright Generated code for initial \rname{sync} rule.}
\label{fig:original_sync_generation}
\end{subfigure}
\begin{subfigure}[b]{0.49\linewidth}
  \begin{lstlisting}[language=C++, keepspaces=true]
// New QUERY rule
h_p.enqueue(SYNC);
// SYNC rule
h_p.sync();
// New QUERY rule
result = f();
  \end{lstlisting}
\caption{Generated code for modified \rname{sync} rule.}
\label{fig:modified_sync_generation}
\end{subfigure}
  \caption{Executing a query $f$}
  \label{fig:compile_query_call}
\end{figure}
the old rule first generates the call,
sends it to the handler,
and then synchronizes (\fig~\ref{fig:original_sync_generation}),
these actions come from the combination of the
\rname{query} and \rname{sync} rule.
The new rule just performs the call after synchronization
occurs (\fig~\ref{fig:modified_sync_generation}).
This approach offers three main benefits:
\begin{itemize}
\setlength{\itemsep}{0.3ex}
\item there is no memory allocation required,
\item no encoding/decoding of the call is required, and
\item which call is being made is known statically.
\end{itemize}
The last item is important,
as now the underlying optimizer knows 
which call is being made, statically.
This allows optimizations such as inlining.

One last optimization uses the knowledge that when the handler
finishes synchronizing with a client, it will have no more work to do.
Therefore, it control passes directly from the handler to the client,
using the scheduling layer of the lightweight threads to avoid
global scheduler.
This optimization is safe, because the handler will otherwise just be idle, and avoids unnecessary context switching.

\subsection{Multi-reservation Separate Blocks}
The code generation for the
multi-reservation separate block differs
slightly from the single-reservation case which
is optimized due to it being a simpler operation.
One can see in \figref{compiled_multi_separate} that
\begin{figure}[htb]
  \footnotesize
  \centering
\begin{centerbox}
  \begin{lstlisting}[language=C++]
client.new_reservations ();
client.add_handler (h1);
client.add_handler (h2);
client.reserve_handlers();

private_queue* h1_p = client.queue_for (h1);
private_queue* h2_p = client.queue_for (h2);

<compiled body>

h1_p.enqueue (END);
h2_p.enqueue (END);
  \end{lstlisting}
\end{centerbox}
\vspace{0.5ex}
\caption{A compiled 2-reservation \lstinline{separate} block}
  \figlabel{compiled_multi_separate}
\end{figure}
some of the complexity is pushed into the client run-time library.
The run-time maintains structures that allow the multiple handlers
to be stored.
The interface between the compiled code and run-time consists of
marking the start of a new set of reservations with
\lstinlinec{new_reservations},
adding a handler with \lstinlinec{add_handler},
and finally safely reserving all handlers with \lstinlinec{reserve_handlers}.
The client can now retrieve the private queues that
were just reserved;
they do not need to be inserted into the handler's
queue-of-queues because the reservation mechanism has already done that.
Signalling the end of the private queue is done as before.
Currently, the multiple reservation implementation
uses one spinlock for every handler to maintain the ordering guarantees.
However, since the number of memory accesses to enqueue
in the queue-of-queues is quite small,
a more sophisticated implementation could use transactional memory
to implement the same behaviour.
These spinlocks were not found to decrease performance.

%%% Local Variables:
%%% mode: latex
%%% TeX-master: main.tex
%%% End:

% \iftoggle{LONGVERSION}
% {
%   \input{implementation_long}
% }
% {
%   \input{implementation_short}
% }

\iftoggle{LONGVERSION}
{
  
\subsection{Removing Redundant Synchronization}
The SCOOP model essentially prevents data races by mandating that
one must access (read and write) separate areas of memory through their respective
handlers.
Due to this, a common SCOOP idiom is that memory is often
copied back and forth between processors when a local
copy is desired for speed reasons.
One example of this is sending data to a worker
for further asynchronous processing.
When copying data in SCOOP there are essentially two options:
push or pull.
Either the data is copied via routines that
asynchronously push data to a \eif{separate} target,
or the data is synchronously pulled by the client that wants it using queries.
Even though the first option appears to enable more concurrency
because it is asynchronous, it often isn't the case.
Consider sending an array one integer at a time:
this involves reading the integer from the client,
packaging the call that will set the integer on the handler,
sending the call,
then applying the call.
The speed advantage of utilizing more than one core is
dwarfed by the huge cost of issuing the call.
Also, the second option (synchronous pull) tends to be more natural,
as the client knows how and where to reconstruct the data.

Therefore it is natural to make queries as efficient as possible.
This was partially addressed using the approach in the previous section,
using sync operations and executing the query on the client.
There is a further enhancement that can be made to this approach,
which is eliding unnecessary sync calls.
A sync call is not necessary if the previous call to the desired
handler was also a sync call;
basically if the handler is already ``synced'' it doesn't need to be
re``synced''.

We perform this elision in two ways:
either by dynamically recording
the synced status in the runtime and ignoring sync operations
on handlers that have already been synced,
or statically by performing a static analysis.

\subsubsection{Dynamic Avoidance}\label{sec:dynamic_avoidance} 
The dynamic method keeps the synced status in the
private queue structure.
When a sync call is made on a private queue,
nothing happens if the queue is already synchronized;
the call merely returns
and the synced status is unaffected.
If the queue is not currently synced,
the sync message is sent to the handler
as usual and when it returns
the synced flag is set in the handler
reflecting that the handler
is processing this private queue,
but the queue is empty.

\subsubsection{Static Removal}\label{sec:static_removal}
The static analysis starts by traversing the control flow graph (CFG).
It annotates every basic block,
basic blocks being sequences of basic instructions,
with a set of handlers that are synchronized by the end of the
block.
This set of handlers is called the \emph{sync-set}.
The traversal of a function's basic blocks can be seen in
\fig~\ref{fig:sync_set_calc_function}.
\begin{figure}[htb]
  \centering
  \footnotesize
  \begin{lstlisting}[language=C++, basicstyle=\sffamily]
while changed $\neq \emptyset$
    b $\in$ changed, changed := changed $-\ \{\mathtt{b}\}$
    common := $\bigcap$ b.predecessors.sync_set

    if b.sync_set $\neq$ UpdateSync (b, common)
        b.sync_set := UpdateSync (b, common)
        changed := changed $\cup$ b.successors
\end{lstlisting}  
  \caption{Sync-set calculation for a function}
  \label{fig:sync_set_calc_function}
\end{figure}
Each block acts as a sync-set transformer, adding and removing
handlers from the set.
As an initial input,
the intersection of the sync-sets of all the block's predecessors
is used.
The traversal continues until every basic block's sync-set has
stopped changing.

Of course this only says how the blocks are traversed,
not how a given block's sync set is calculated given the
instruction in that block.
This is described in the \lstinline[basicstyle=\sffamily]{UpdateSync} function,
shown in ~\fig~\ref{fig:sync_set_calc}.
\begin{figure}[htb]
  \centering
  \footnotesize
  \begin{lstlisting}[language=C++, escapechar=^, basicstyle=\sffamily]
UpdateSync (b, synced):
    for  inst $\in$ b
        h := HandlerOf(inst)
        synced :=
        ^\vskip-3.05em\[\phantom{xx}\begin{array}{ll}
\mbox{synced} \cup \{\mbox{h}\} &
       \text{\textrm{if \textsf{inst} is sync.}} \\
\mbox{synced} - \{\mbox{h}\} &
       \text{\textrm{if \textsf{inst} is async.}} \\
\emptyset &
       \text{\textrm{if \textsf{inst} has side effects}} \\
\mbox{synced} & \; \text{\textrm{otherwise}}
\end{array}
\]
\vskip-0.8em
^

    return synced

  \end{lstlisting}
  \caption{Sync-set calculation for a block}
  \label{fig:sync_set_calc}
\end{figure}
%
    % if (is_sync (inst)))
    %   synced := )
    % else if ((h_p = is_async (&inst)))
    %   synced = clear_may_alias (h_p, synced);
    % else if (CallInst *call =
    %          dyn_cast<CallInst>(&inst))
    %   synced = clear_synced_for_call(call, synced);
Each type of instruction is handled differently:
synchronization calls add the target handler to the sync-set,
asynchronous calls remove those handlers (and anything they may be aliased to),
and arbitrary calls clear the sync-set entirely.
Obviously this final case is quite severe,
as it has to be, because a call could subsequently
issue asynchronous calls on all the handlers currently in the sync-set.
This can be mitigated by not clearing the sync-set for functions
which are marked with the \texttt{readonly} and \texttt{readnone} flags.
LLVM will automatically add these flags when it can determine
that they hold.

The static analysis operates on LLVM bitcode,
and is implemented as a standard LLVM pass (outside of the base compiler).
Keeping the pass outside of the base \qsname compiler
has the advantage that it separates the generation of code from
the analysis and transformation of the generated control flow graph.

\newcommand{\qssync}[1]{\lstinline{#1.sync()}}

\tikzset{
  basic block/.style={rectangle, draw, thick, rounded corners,
    top color=white, bottom color=#1!50!black!20, draw=#1!50!black!20,
    drop shadow={opacity=1.0, fill=gray!30!white}}
}

\newcommand{\looptikz}[4]{
  \footnotesize
\begin{tikzpicture}[scale=0.9]
  \clip (-8em,-12em) rectangle (7em, 12em);
  \node (loop) [basic block=blue, draw, fill=blue!5, align=left,
                minimum height=2cm, minimum width=2cm]
       {#1};
  \node (pre_header) [basic block=blue, fill=blue!5, above=1cm of loop,
                      minimum height=1cm, minimum width=2cm]
                      {\qssync{h\_p}};

  \draw [draw, -stealth]
       (pre_header.south) to node [right] {\{#3\}} (loop.north) {};
 
  \node [basic block=green, minimum height=0.7cm, minimum width=1cm,
         anchor=west] (true_br) at (loop.south west) {True};
  \node [basic block=red, fill=red!20, minimum height=0.7cm, minimum width=1cm,
         anchor=east] (false_br) at (loop.south east) {False};

  \draw [-stealth]
    (true_br.south) to[out=215, in=135, looseness=3]
       node [right] {\{#2\}} ($(loop.north)+(-1.5em,0)$);

  \node (exit) [basic block=blue, minimum height=1cm, minimum width=2cm,
                below=1cm of loop] {#4};

  \draw[-stealth]
       (false_br.south) to[out=270, in=90] node [right=0.5em, midway] {\{#2\}} (exit.north) {};

  \node (b1) [right=0.1cm of pre_header] {B1};
  \node (b2) [right=0.1cm of loop] {B2};
  \node (b3) [right=0.1cm of exit] {B3};
%\draw [brown] (current bounding box.south west) rectangle (current bounding box.north east);
\end{tikzpicture}
}

\subsubsection{Example}
The effect of the sync coalescing pass can be seen
in \fig~\ref{fig:loop_sync}.
\begin{figure}[htb]
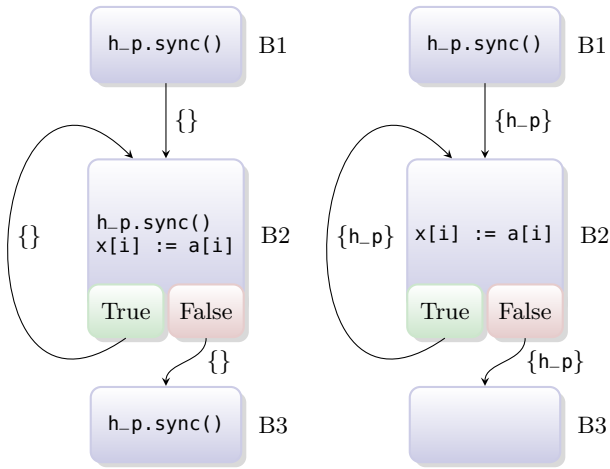

\vspace{-4ex}
  \centering
\begin{subfigure}[b]{0.49\linewidth}
%\frame{
  \looptikz{\qssync{h\_p}\\\lstinline{x[i] := a[i]}}{}{}{\qssync{h\_p}}
%}
\vspace{-3ex}
\caption{A simple loop before the sync-coalescing pass.}
\label{fig:before_pass}
\end{subfigure}
\begin{subfigure}[b]{0.49\linewidth}
%\frame{
  \looptikz{\lstinline{x[i] := a[i]}}{\lstinline{h\_p}}{\lstinline{h\_p}}{}
%}
\vspace{-3ex}
\caption{After sync-coalescing sync-sets label edges.}
\label{fig:after_pass}
\end{subfigure}
\caption{Sync-coalescing pass}\label{fig:loop_sync}
\end{figure}
This program has three blocks,
with sync operations in each one.
Before the sync-coalescing pass, in \fig~\ref{fig:before_pass},
the client is reading values out of a handler's array,
for which a na\"{i}ve code generator will produce a sync before every array read.
\fig~\ref{fig:after_pass} shows the results of the sync-coalescing pass
in such a situation.
The sync-sets are shown explicitly on the edges out of each block.
In this case there are no calls that may invalidate a sync-set,
so the handler \lstinline{h_p} appears on all edges.
The result of this is that the sync calls in blocks B2 and B3 can be removed.
Removing sync calls in the body of a loop can greatly increase performance.
Note that even though the sync call in the body of B2 was removed,
\lstinline{h_p} still appears on B2's outgoing edges because B2 doesn't invalidate
the synchronization on \lstinline{h_p} by issuing an asynchronous call.

It is not always possible, however,
to remove the sync operations,
even when the processor is apparently unaffected.
Consider \fig~\ref{fig:failed_loop_sync},
\begin{figure}[htb]
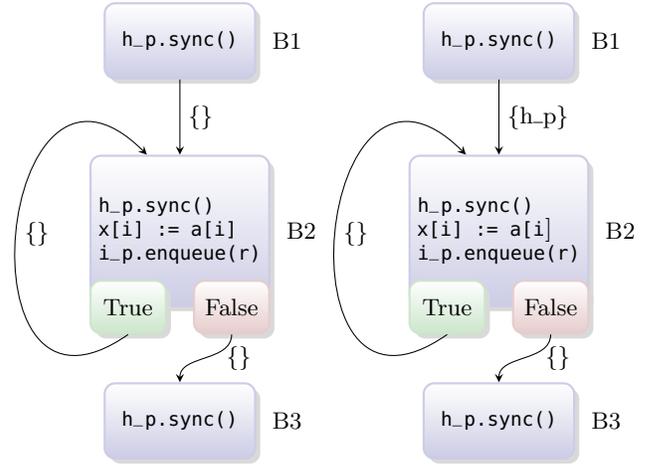

\vspace{-2ex}
  \centering
\begin{subfigure}[b]{0.49\linewidth}
%\frame{
  \looptikz{\qssync{h\_p}\\\lstinline{x[i] := a[i]}\\\lstinline{i\_p.enqueue(r)}}{}{}{\qssync{h\_p}}
%}
\vspace{-3ex}
\caption{A simple loop with an extra asynchronous call on $i$.}
\label{fig:before_pass_failed}
\end{subfigure}
\begin{subfigure}[b]{0.49\linewidth}
%\frame{
  \looptikz{\qssync{h\_p}\\\lstinline{x[i] := a[i}]\\\lstinline{i\_p.enqueue(r)}}{}{h\_p}{\qssync{h\_p}}
%}
\vspace{-3ex}
\caption{Handlers \lstinline{h_p} and \lstinline{i_p} may be aliased: no coalescing.}
\label{fig:after_pass_failed}
\end{subfigure}
\caption{Ineffective sync-coalescing pass }\label{fig:failed_loop_sync}
\end{figure}
where there is an additional call to \lstinline[language=C++]{i_p.enqueue(r)},
in \fig~\ref{fig:before_pass_failed}.
Enqueueing a call is an asynchronous activity,
but it occurs on a different handler variable.
This is not enough, though, to conclude the handler \lstinline{h_p} is
unaffected,
as these are only variables and
could be aliased to one another.
Meaning they effectively be the same handler.
This means that at the end of the B2 block the outgoing edges are labeled,
visible in in \fig~\ref{fig:after_pass_failed},
with neither \lstinline{h_p} or \lstinline{i_p}.
If more aliasing information is given to the compiler
then it is possible that this ambiguity can be resolved and
\lstinline{h_p} can be added to the sync-set for the block.

The static analysis is important as it goes further towards
getting SCOOP out of the way of the optimization passes.
In the end our implementation uses both the static and dynamic approaches.
The static analysis is used when it can be,
but it is necessarily conservative.
For the cases where the static analysis keeps
an unnecessary sync operation around,
the dynamic check will eliminate the round-trip to the handler.

}
{
  \input{remove_sync_short}
}

%}}}

%{{{ Optimization evaluation
\section{Optimization Evaluation}\label{sec:optimization_evaluation}

\iftoggle{LONGVERSION}
{
  
Here we examine the impact of the following optimizations (also outlined in 
Section~\ref{sec:implementation}):
\begin{itemize}
\setlength{\itemsep}{0.3ex}
\item Applying no optimizations (\textbf{None}).
\item Dynamically coalescing sync operations by recording and checking
  the synchronization status in the runtime (\textbf{Dynamic}),
  as in Section~\ref{sec:dynamic_avoidance}.
\item Statically determining unnecessary sync operations 
  and removing them in a compiler pass (\textbf{Static}),
  as in Section~\ref{sec:static_removal}.
\item Usage of the queue-of-queues and private queues as a
  handler/client communication abstraction (\textbf{QoQ}),
  as seen in the semantic model given in Section~\ref{sec:model}.
\item Applying all optimizations (\textbf{All}).
\end{itemize}
For the comparison, a variety of workloads are used.

\subsection{Computation and Coordination-based Workloads}
Properly evaluating a core runtime mechanism,
such as {\qsname},
requires that it be used in a diverse assortment of situations.
With this in mind, we categorize the benchmarks
that will be used into two main groups:
\begin{itemize}
\setlength{\itemsep}{0.3ex}
\item \emph{parallel}:
  problems where concurrency is not part of the
  functional specification, but can be used to speed up the execution.
\item \emph{concurrent}:
  problems which are defined by their concurrent behaviour.
\end{itemize}
The first work type, parallel, is often a data processing task,
where multiple threads each process part of a large data set to
decrease the total running time.
The second type of work, concurrent,
is more about the coordination between the threads of control.
Here, the concurrency is part of the system's specification,
such as a server handling multiple clients simultaneously.

\subsubsection{Parallel Workloads}
The benchmark programs we selected for the parallel problems
are a selection from the Cowichan problem set~\cite{Wilson95assessingand}.
They focus on numerical processing and working over large
arrays and matrices.
The programs include:
\begin{itemize}
\setlength{\itemsep}{0.3ex}
\item \bname{randmat}: randomly generate a matrix of size $n_r$.
\item \bname{thresh}: pick the top $p$\% of a matrix of size $n_r$
  and construct a mask.
\item \bname{winnow}: apply a mask to a matrix of size $n_r$,
  sorting the elements that passed the mask based on their value and position,
  and taking only $n_w$ from that sorted list.
\item \bname{outer}:
  constructing a matrix and vector based off a list of points.
\item \bname{product}: matrix-vector product.
\end{itemize}
These benchmarks can be sequentially composed together,
the output of one becoming the input to the next,
to form a \bname{chain}.
This chain is more complex and sizable than the individual
and gives a more diverse picture of a language's parallel performance.

\subsubsection{Concurrent Workloads}
The concurrent problems focus on the interaction of different independent
threads with each other.
We have created three benchmarks that represent different interaction
patterns:
\begin{itemize}
\setlength{\itemsep}{0.3ex}
\item \bname{mutex}: $n$ threads all compete for access to a single resource,
  the threads do not depend on each other.
\item \bname{prodcons}: $n$ producers and $n$ consumers each operate on
  a shared queue; the queue has no upper limit so producers do not depend
  on consumers, but consumers must wait until the queue is non-empty to
  make progress.
\item \bname{condition}: $n$ ``odd'' and $n$ ``even'' workers
  increment a variable from an odd (even) to an even (odd) number.
  Each group depends on the other to make progress.
\end{itemize}
All of the above are repeated for $m$ iterations.
Finally to this we add two concurrency benchmarks from the
Computer Language Benchmarks Game~\cite{benchmarksgame}:
\begin{itemize}
\setlength{\itemsep}{0.3ex}
\item \bname{threadring}:
  threads pass a token around a ring in sequence until the token
  has been passed $n_t$  times.
\item \bname{chameneos}:
  colour changing ``chameneos'' mate and change their colours depending
  on who they mate with. This is done $n_c$ times.
\end{itemize}
The combination of these parallel and concurrent benchmarks
gives us a balanced view of the performance characteristics of the
approach.

\paragraph{Setup} All benchmarks were performed 20 times on a
Intel Xeon Processor E7-4830 server
(4 $\times$ 2.13 GHz, each with 8 cores; 32 physical cores total) with
256 GB of RAM,
running Red Hat Enterprise Linux Server release 6.3.
Language and compiler versions used were:
gcc-4.8.1, go-1.1.2, ghc-7.6.3, erlang-R16B01.
For the parallel benchmarks,
the problem sizes used are
$n_r = \text{10,000}$,
$p = \text{1}$ and
$n_w = \text{10,000}$;
for the concurrent benchmarks
$n = \text{32}$,
$m = \text{20,000}$,
$n_t = \text{600,000}$, and
$n_c = \text{5,000,000}$.

\subsection{Parallel Benchmarks}
As mentioned in the previous section,
the idiomatic way to transfer data in {\qsname} is
to have the client pull data from the handler.
This happens in the Cowichan problems often,
as the underlying data structures are almost exclusively large arrays
and they must be distributed to and from workers.
This is important for interpreting the results of the comparison
of optimizations when they are applied to the Cowichan problems.
\fig~\ref{fig:optimization_compare_parallel}
\begin{figure*}
  \centering
  \includegraphics[width=\linewidth]{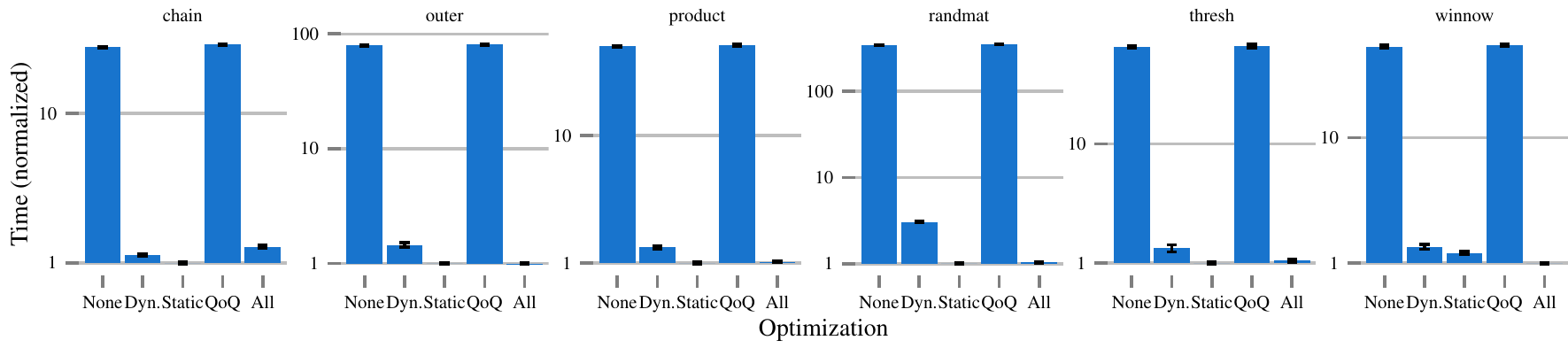}
  \caption{Communication times for different optimization techniques evaluated on parallel tasks }
  \label{fig:optimization_compare_parallel}
\end{figure*}
displays the communication time
(as this is by far the factor dominating the execution time)
normalized to the fastest version and in log scale.
From this, it is clear that there is
a marked improvement in communication time for
the parallel benchmarks over
the programs that had no optimizations that
reduce the number of sync operations that are performed.

The difference between having no reduction in
the number of sync calls (\textbf{None}, \textbf{QoQ}) and
employing some reduction technique
(\textbf{Dynamic}, \textbf{Static}, \textbf{All})
is that the latter is at least 10 times faster (\bname{chain}),
near 100 times faster
(\bname{thresh}, \bname{winnow}, \bname{outer}, \bname{product}),
or over 100 times faster (\bname{randmat}).

Reducing the number
of sync calls is a great benefit for programs
that require large numbers of synchronous calls.
There is an additional property of these benchmarks that they access
the arrays in a very regular way, namely they copy from one array to another,
generally calling sync many times in a tight loop.
In these cases,
it is beneficial to fully lift this call right out of the loop body,
as the \textbf{Static} optimization does.
The difference between the \textbf{Static} and \textbf{Dynamic}
sync-coalescing optimizations can be seen in more detail in
Table~\ref{tab:parallel_optimization}.
We can see for example that randmat is 3x faster when using the \textbf{Static}
sync coalescing optimization compared to the \textbf{Dynamic},
and others using \textbf{Static} sync coalescing are between 1.1x and 1.4x faster
than \textbf{Dynamic}.
\begin{table}
\caption{Normalized (to fastest) comparison of optimizations on parallel tasks}
\label{tab:parallel_optimization}
\centering
{\footnotesize
\begin{tabular}{lrrrrr}
  \hline
  Task    & none   & Dyn. & Static & QoQ    & All  \\ 
  \hline
  chain   & 27.70  & 1.13 & 1.00   & 28.81  & 1.28 \\ 
  outer   & 78.95  & 1.45 & 1.00   & 80.44  & 1.00 \\ 
  product & 49.99  & 1.33 & 1.00   & 51.18  & 1.02 \\ 
  randmat & 345.61 & 3.05 & 1.00   & 353.43 & 1.03 \\ 
  thresh  & 64.54  & 1.33 & 1.00   & 66.08  & 1.05 \\ 
  winnow  & 53.14  & 1.35 & 1.21   & 54.33  & 1.00 \\ 
  \hline
\end{tabular}
}
\end{table}

% \lstdefinelanguage
% [x64]{Assembler}
% [x86masm]{Assembler}
% {morekeywords={vmovdqu, vpmulld, r12, r14, r15, rcx, xmm0, xmm1, xmm2}}

% \lstset{language = [x64]Assembler}

% \begin{lstlisting}
%   vmovdqu	(%r15,%rcx,4), %xmm0
%   vmovdqu	16(%r15,%rcx,4), %xmm1
%   vmovdqu	(%r12,%rcx,4), %xmm2
%   vpmulld	%xmm0, %xmm2, %xmm0
%   vpmulld	16(%r12,%rcx,4), %xmm1, %xmm1
%   vmovdqu	%xmm0, (%r14,%rcx,4)
%   vmovdqu	%xmm1, 16(%r14,%rcx,4)
% \end{lstlisting}

\subsection{Concurrent Benchmarks}
While the \textbf{QoQ} optimization 
has little effect on the Cowichan problems,
it has a stronger influence on the concurrency problems 
which can be seen in \fig~\ref{fig:optimization_compare_concurrent}.
\begin{figure*}[htb]
  \centering
  \includegraphics[width=\linewidth]{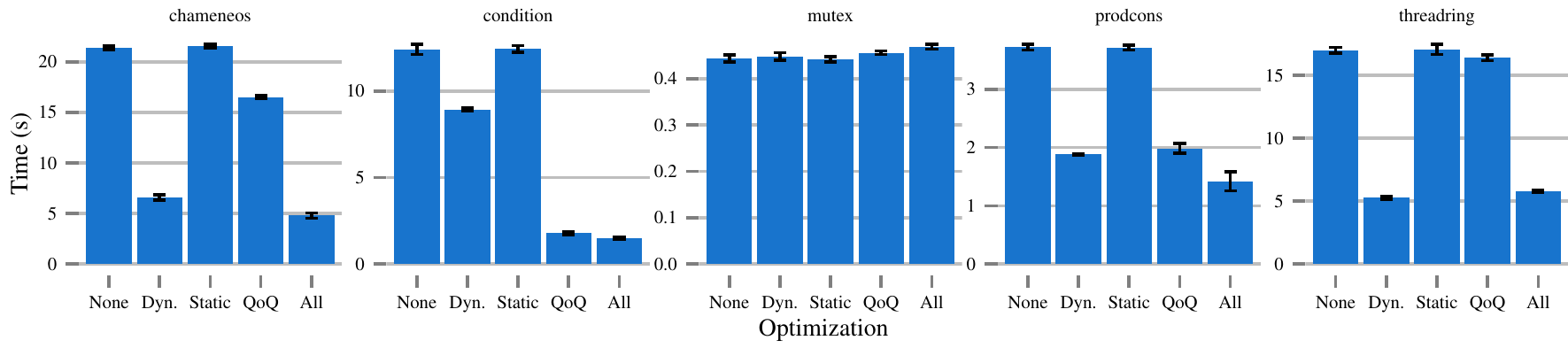}
  \caption{Comparison of {\qsname} optimizations on concurrent benchmarks}
  \label{fig:optimization_compare_concurrent}
\end{figure*}
The full results are visible in Table~\ref{tab:concurrent_optimization}.
\begin{table}
\caption{Times (in seconds) for optimizations applied on concurrent benchmarks}
\label{tab:concurrent_optimization}
\centering
{\footnotesize
\begin{tabular}{lrrrrr}
  \hline
  Task       & none  & Dyn. & Static & QoQ   & All  \\ 
  \hline
  chameneos  & 21.41 & 6.58 & 21.58  & 16.54 & 4.80 \\ 
  condition  & 12.41 & 8.93 & 12.44  & 1.78  & 1.50 \\ 
  mutex      & 0.44  & 0.45 & 0.44   & 0.46  & 0.47 \\ 
  prodcons   & 3.72  & 1.88 & 3.71   & 1.98  & 1.42 \\ 
  threadring & 17.01 & 5.27 & 17.08  & 16.41 & 5.80 \\ 
  \hline
\end{tabular}
}
\end{table}

Using \textbf{QoQ} is faster (\bname{chamenos, condition, prodcons}) or
about the same (\bname{mutex, threadring}).
The increase in performance can be attributed to the better utilization
of available processing capabilities resulting from
less blocking in the underlying semantics.
When using \textbf{QoQ} there are also fewer context switches,
since the private queues require only one context switch to wait for
a query to return.
When not using the queue-of-queues a client must wait three times:
first for the lock on the handler,
then the handler must wait for the client to log its query,
then the client must wait again for the handler to return the query.
However, the queue-of-queues approach does introduce some overhead
in cases where there are few calls issued in the \lstinline{separate} block,
because it must enqueue twice,
once putting the private queue into the queue-of-queues,
then putting the call into the private queue.
However, mostly the \textbf{QoQ} optimization is beneficial.

One can also see here the importance of
the \textbf{Dynamic} sync-coalescing optimization.
In particular it benefits
the \bname{chameneos, condition, prodcons} and \bname{threadring} benchmarks.
This underscores the benefit of limiting the roundtrips to the handler
when they are not necessary.
When it applies, this increases the performance between
1.5 and 3$\times$.
It is important to note here that because the workloads are irregular,
the \textbf{Static} sync-coalescing is not as effective.
Its main benefit comes in cases of regular access patterns where
it can be applied more readily.

\subsection{Summary}
Each optimization has particular situations in which it brings
the most benefit:
\begin{itemize}
\setlength{\itemsep}{0.3ex}
\item \textbf{QoQ} is best on coordination tasks but is not as
  useful for query-heavy workloads.
\item \textbf{Dynamic} sync-coalescing is useful on
  both coordination tasks and tasks with many queries.
\item \textbf{Static} sync-coalescing is primarily effective
  on very structured query usages, beating even \textbf{Dynamic}
  in such situations.
\end{itemize}
The geometric mean of all benchmarks
is 20.70s for no optimizations,
1.99s for \textbf{Dynamic} sync-coalescing,
2.24s for \textbf{Static} sync-coalescing,
16.21s for \textbf{QoQ},
and 1.36s with all optimizations.
The net effect is that the final {\qsname} runtime is
\smallsim 15$\times$ faster than the basic runtime.

\subsection{Application outside the Prototype}
To provide further evidence of the effectiveness of
the execution techniques presented here,
a new runtime was constructed for the research branch, EVE\cite{se:2014:eve}, of the EiffelStudio IDE.
The new runtime, which we've named EVE/Qs,
incorporates the \textbf{QoQ} and \textbf{Dynamic} optimizations.
The \textbf{Static} optimization was not implemented due to the lack of
robust static code analysis and transformation facilities in EiffelStudio.
The results are promising:
the speedup compared to the existing production SCOOP runtime has a geometric mean of
11.7$\times$ on the concurrency benchmarks,
7.7$\times$ on the parallel benchmarks,
and a 9.7$\times$ across all benchmarks.
The absolute performance of EVE/Qs is lower compared to {\qsname} because
EVE/Qs inherits several implementation decisions from EiffelStudio.
For example, the use of a shadow-stack for garbage collection,
inhibiting efficient tight-loop optimizations important for the parallel benchmarks.
Also, storing handler IDs in the object header requiring maintenance of a
secondary thread-safe data structure to lookup the handler data.
Since the handlers are accessed incredibly often,
this also is a detriment to performance.
Lastly, since {\qsname} uses lightweight threads,
situations with high contention, such as in the concurrency benchmarks, benefit.

}
{
  \input{optimization_eval_short}
}

%}}}

%{{{ Language Comparison
\section{Language Comparison}\label{sec:language_comparison}

\iftoggle{LONGVERSION}
{
  It is difficult to gauge the appropriateness of a
concurrency model without comparing it against its contemporaries.
With that in mind, this section presents a comparison of
{\qsname} with four well-established languages.

\subsection{A Variety of Languages}
The comparison languages should:
be modern, well-known, and represent a variety of different
underlying design choices.
For the purposes of the evaluation,
this means that we should select from different programming paradigms,
different approaches to shared memory,
different concurrency safety guarantees,
and different threading implementations.
We have chosen a selection of languages:
C++/TBB (Threading Building Blocks)~\cite{5672517},
Erlang~\cite{Armstrong:1996:CPE:229883},
Go~\cite{golang}, and
Haskell~\cite{PeytonJones:1996:CH:237721.237794}. 
This selection attempts to combine a reasonable number of the facets
outlined above to give a complete picture.
To make the diversity clear, we present this in
Table~\ref{tab:language_characteristics}.
\begin{table*}[htb]
\caption{Language characteristics}
\label{tab:language_characteristics}
\footnotesize
\centering
\begin{tabular}{cccccc}
  \hline
  Language & Races    & Threads & Paradigm   & Memory     & Approach \\ \hline
  C++/TBB  & possible & OS      & Imperative & Shared     & Skeletons/traditional \\
  Go       & possible & light   & Imperative & Shared     & Goroutines/channels \\
  Haskell  & none     & light   & Functional & STM        & STM/Repa \\
  Erlang   & none     & light   & Functional & Non-shared & Actors \\
  {\qsname}& none     & light   & O-O        & Non-shared & Active Objects \\
  \hline
\end{tabular}
\end{table*}

The Memory column refers to how memory is shared between threads.
Erlang has no sharing between different processes:
when data is sent between processes
it is copied in its entirety.
In {\qsname} the programmer is only able ti access shared memory
through a handler.
In Haskell it is perfectly possible to construct data races if
one uses mutable references.
However, to include one more race-free model to this comparison,
we restrict ourselves to using Haskell's STM implementation,
the \texttt{par} construct which executes pure computations in parallel,
as well as the Repa library which is specialized to work on parallel arrays. 

% \edcomm{SW}{Cite Repa paper}

\subsection{Parallel Benchmarks}

The parallel benchmarks are meant to measure
how well a language can handle taking a particular
program and scaling it given more computational
resources (cores).
Note that it is common in the Erlang and {\qsname} implementations of the
Cowichan problems that a
significant amount of time is spent
sharing results among the threads.
Therefore, to more clearly see the effect of different optimizations,
and to separate computational effects from communication effects,
we distinguish the time spent computing versus the time spent
communicating the results.

\subsubsection{Execution Time}
We can see the graph of performance given 32 cores in
\fig~\ref{fig:parallel_benchmarks}.
\begin{figure*}[htb]
  \centering
  \includegraphics[width=\linewidth]{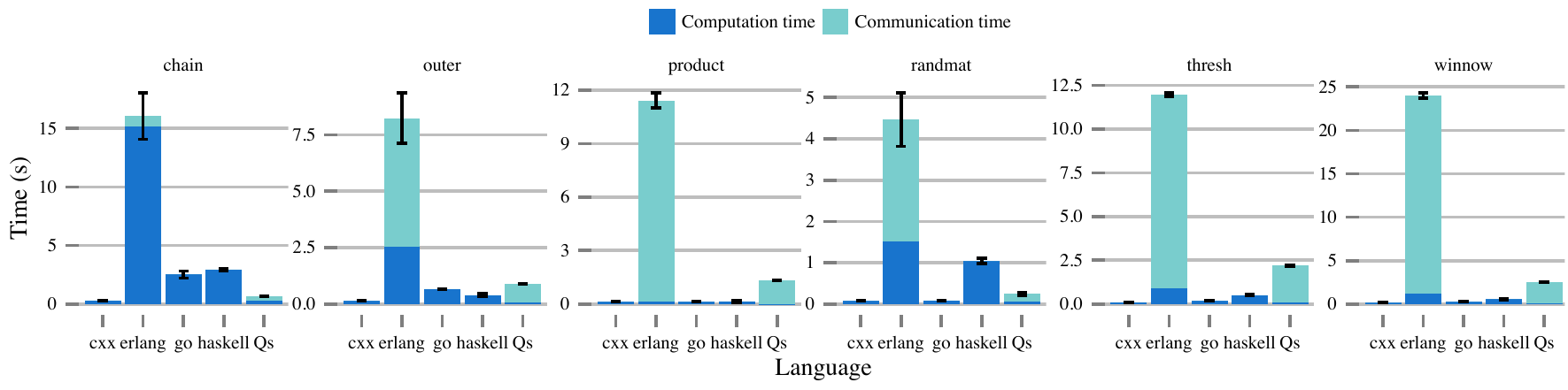}
  \caption{Execution times of parallel tasks on different languages, executed on 32 cores}
  \label{fig:parallel_benchmarks}
\end{figure*}
As with {\qsname}, to give a clearer picture of
the performance characteristics of Erlang,
we also distinguish the computation
time from the communication time.
We can see that {\qsname} and Erlang both spend a majority
of their time in communication,
with the exception of the \bname{chain} problem, 
which has much less communication between the workers.
It is useful to consider both the total and the computation
time: in non-benchmark style problems it is more likely
that the workloads fall somewhere in the middle.
For example, the \bname{chain} problem, which is 
the composition of the other smaller benchmarks,
does not suffer from nearly the same communication
burden as they do.

Erlang has unfavorable performance results compared to
the other languages.
Due to Erlang's data representation
(forces to use linked lists to represent matrices)
and its execution model
(cannot use the HiPE-optimized builds~\cite{erlanghipe} with the multithreaded runtime),
it generally falls far behind the other approaches,
even Haskell and {\qsname}.

\begin{sloppypar}
Besides Erlang,
the other languages are more closely grouped.
The geometric means for total time are,
in increasing order:
C++/TBB (0.32s), Go (0.57s), Haskell (0.89s), {\qsname} (1.35s), 
and Erlang (18.07s).
For computation-only time, the order is:
{\qsname} (0.29s), C++/TBB (0.32s), Go (0.57s), Haskell (0.89s) and
Erlang (4.32s).
\end{sloppypar}

Note that this puts {\qsname} first because many of the cache effects
are removed due to the predistribution of the data before
the timing starts.
This is only included as a sanity test,
to show that the lower-bound for the {\qsname} implementation
is competitive with the other approaches.

\subsubsection{Scalability} The other aspect that we investigated was the speedup of
the benchmarks across 32 cores.
In \fig~\ref{fig:parallel_speedup}
\begin{figure*}[htb]
  \centering
  \includegraphics[width=\linewidth]{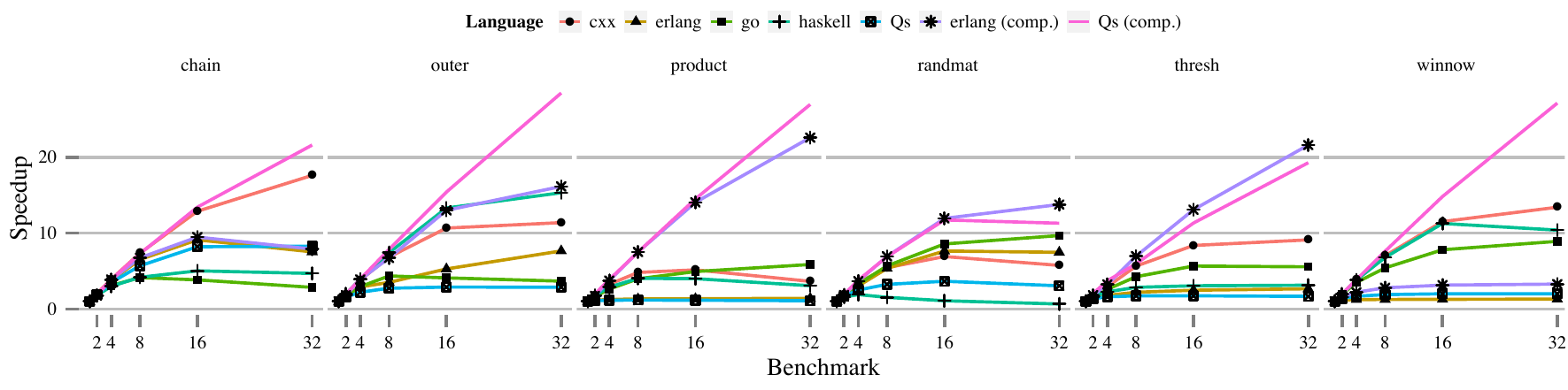}
  \caption{Speedup over single-core performance, up to 32 cores}
  \label{fig:parallel_speedup}
\end{figure*}
we can see the performance of
the various languages on the different problems.
We can see that on \bname{chain}, 
most languages manage to achieve a speedup of at least 5x.
Go is the exception to this, and performance decreases past 8 cores.
Erlang also sees a performance degradation,
though only past 16 cores.
This was also an effect that was seen in 
our language comparison
study~\cite{nanz-et-al:2013:benchmarking-multicore-languages}
from which the implementation was taken;
the implementation was also reviewed by a key Go developer in the study.

Also of note is the performance of Haskell on the \bname{randmat} benchmark.
This is one of the few benchmarks where Repa could not be effectively
used due to the nature of the problem,
so the basic \texttt{par}-based concurrency primitives are used.
The basic strategy has chunks of the output array constructed
in parallel, then concatenated together.
The concatenation is sequential, however, putting a limit on the
maximum speedup;
using the ThreadScope~\cite{Wheeler:2010:VMM:1673012.1673015} performance 
reporting tool, we could see that
the stop-the-world garbage collector was intervening too often.
The last unexpected result was the inability for the Erlang
version of the \bname{winnow} program to speedup past about 2-3x.
This was examined in detail but no cause could be found.
Precise timing data can be found in Table~\ref{tab:parallel}.

\newcommand{\perfcolwidth}{0.75cm}
\newcommand{\pcw}{p{\perfcolwidth}}
\newcommand{\perfcolspec}{p{1.1cm} p{1.6cm} p{0.3cm} \pcw  \pcw \pcw  \pcw \pcw  \pcw}
\begin{table*}[htb]
\caption{Parallel benchmark times (in seconds)}
\label{tab:parallel}
{

% this table is tight, cut down on the whitespace
\renewcommand{\tabcolsep}{2pt}

\footnotesize
\parbox{0.4999\linewidth}{
\centering
\begin{tabular}{\perfcolspec}%{lllrrrrrr}
  \hline
         &           &   & \multicolumn{6}{c}{Threads}                   \\
  Task   & Lang      & V & 1     & 2     & 4     & 8     & 16    & 32    \\ 
  \hline
 randmat & c++       & T & 0.44  & 0.23  & 0.13  & 0.08  & 0.06  & 0.08  \\ 
 randmat & erlang    & T & 30.93 & 18.01 & 10.20 & 5.77  & 4.05  & 4.14  \\ 
 randmat & erlang    & C & 20.69 & 11.26 & 5.63  & 2.99  & 1.73  & 1.50  \\ 
 randmat & go        & T & 0.78  & 0.43  & 0.24  & 0.14  & 0.09  & 0.08  \\ 
 randmat & haskell   & T & 0.68  & 0.43  & 0.36  & 0.44  & 0.62  & 1.03  \\ 
 randmat & {\qsname} & T & 0.72  & 0.43  & 0.29  & 0.22  & 0.21  & 0.23  \\ 
 randmat & {\qsname} & C & 0.59  & 0.30  & 0.15  & 0.08  & 0.05  & 0.05  \\ 
 thresh  & c++       & T & 1.00  & 0.66  & 0.34  & 0.18  & 0.12  & 0.11  \\ 
 thresh  & erlang    & T & 31.82 & 22.35 & 17.77 & 14.48 & 12.88 & 11.96 \\ 
 thresh  & erlang    & C & 19.30 & 10.74 & 5.97  & 2.77  & 1.47  & 0.89  \\ 
 thresh  & go        & T & 0.95  & 0.60  & 0.37  & 0.22  & 0.17  & 0.17  \\ 
 thresh  & haskell   & T & 1.56  & 0.96  & 0.69  & 0.55  & 0.51  & 0.50  \\ 
 thresh  & {\qsname} & T & 3.71  & 2.72  & 2.28  & 2.10  & 2.11  & 2.15  \\ 
 thresh  & {\qsname} & C & 1.87  & 1.08  & 0.54  & 0.31  & 0.16  & 0.09  \\ 
 winnow  & c++       & T & 2.04  & 1.03  & 0.53  & 0.29  & 0.18  & 0.15  \\ 
 winnow  & erlang    & T & 31.03 & 26.02 & 25.04 & 24.75 & 24.38 & 23.95 \\ 
 winnow  & erlang    & C & 4.06  & 2.58  & 1.84  & 1.46  & 1.29  & 1.24  \\ 
 winnow  & go        & T & 2.47  & 1.29  & 0.71  & 0.46  & 0.32  & 0.28  \\ 
 winnow  & haskell   & T & 5.43  & 2.77  & 1.42  & 0.80  & 0.48  & 0.52  \\ 
 winnow  & {\qsname} & T & 5.16  & 3.74  & 3.04  & 2.69  & 2.58  & 2.57  \\ 
 winnow  & {\qsname} & C & 2.83  & 1.40  & 0.72  & 0.36  & 0.19  & 0.10  \\ 
   \hline
\end{tabular}
}
\hspace{0.1ex}
\parbox{0.4999\linewidth}{
\centering
\begin{tabular}{\perfcolspec}%{lllrrrrrr}
  \hline
          &          &   & \multicolumn{6}{c}{Threads}                    \\
  Task   & Lang      & V & 1      & 2     & 4     & 8     & 16    & 32    \\ 
  \hline
 outer   & c++       & T & 1.59   & 0.83  & 0.42  & 0.23  & 0.15  & 0.14  \\ 
 outer   & erlang    & T & 61.57  & 38.21 & 21.19 & 17.57 & 11.67 & 8.05  \\ 
 outer   & erlang    & C & 40.66  & 22.54 & 10.45 & 6.05  & 3.12  & 2.52  \\ 
 outer   & go        & T & 2.47   & 1.44  & 0.84  & 0.57  & 0.60  & 0.67  \\ 
 outer   & haskell   & T & 5.49   & 2.76  & 1.40  & 0.74  & 0.41  & 0.36  \\ 
 outer   & {\qsname} & T & 2.58   & 1.62  & 1.15  & 0.93  & 0.90  & 0.89  \\ 
 outer   & {\qsname} & C & 1.87   & 0.93  & 0.46  & 0.24  & 0.12  & 0.06  \\ 
 product & c++       & T & 0.44   & 0.23  & 0.13  & 0.09  & 0.08  & 0.12  \\ 
 product & erlang    & T & 15.89  & 13.94 & 12.66 & 12.08 & 11.82 & 11.33 \\ 
 product & erlang    & C & 3.35   & 1.95  & 0.90  & 0.45  & 0.24  & 0.15  \\ 
 product & go        & T & 0.76   & 0.46  & 0.29  & 0.19  & 0.15  & 0.13  \\ 
 product & haskell   & T & 0.45   & 0.25  & 0.16  & 0.11  & 0.11  & 0.15  \\ 
 product & {\qsname} & T & 1.49   & 1.33  & 1.27  & 1.24  & 1.28  & 1.34  \\ 
 product & {\qsname} & C & 0.32   & 0.16  & 0.08  & 0.04  & 0.02  & 0.01  \\ 
 chain   & c++       & T & 5.57   & 2.76  & 1.42  & 0.76  & 0.43  & 0.32  \\ 
 chain   & erlang    & T & 120.59 & 69.00 & 32.06 & 18.48 & 13.23 & 16.01 \\ 
 chain   & erlang    & C & 119.68 & 68.13 & 30.93 & 17.75 & 12.63 & 15.15 \\ 
 chain   & go        & T & 7.39   & 4.09  & 2.39  & 1.79  & 1.93  & 2.60  \\ 
 chain   & haskell   & T & 13.78  & 7.71  & 4.62  & 3.30  & 2.74  & 2.94  \\ 
 chain   & {\qsname} & T & 5.60   & 2.88  & 1.56  & 0.97  & 0.68  & 0.67  \\ 
 chain   & {\qsname} & C & 5.54   & 2.75  & 1.40  & 0.74  & 0.40  & 0.25  \\ 
   \hline
\end{tabular}
}\\
\begin{center}
 V - Timing Variant, T - Total time, C - Compute time.
\end{center}
}
\end{table*}

\subsection{Concurrent Benchmarks}
The concurrent programming tasks are compared in
\fig~\ref{fig:concurrent_benchmarks} and
\begin{figure*}[htb]
  \centering
  \includegraphics[width=\textwidth]{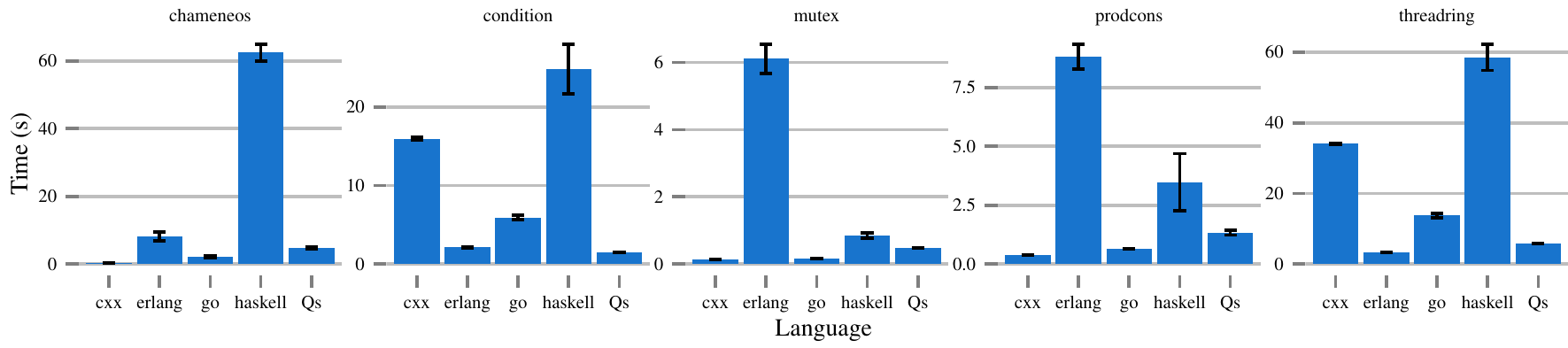}
  \caption{Execution times of concurrent tasks on different languages}
  \label{fig:concurrent_benchmarks}
\end{figure*}
exact times in Table~\ref{tab:concurrent}.
\begin{table}[htb]
\caption{Concurrent benchmark times (in seconds)}
\label{tab:concurrent}
\centering
{
\footnotesize
\begin{tabular}{lrrrrr}
  \hline
      Task  & c++   & erlang & go    & haskell & {\qsname} \\ 
  \hline
 chameneos  & 0.32  & 8.67   & 2.40  & 61.97   & 4.71 \\ 
 condition  & 15.92 & 2.15   & 5.95  & 26.05   & 1.48 \\ 
 mutex      & 0.14  & 6.13   & 0.17  & 0.86    & 0.47 \\ 
 prodcons   & 0.40  & 8.78   & 0.66  & 2.99    & 1.33 \\ 
 threadring & 34.13 & 3.30   & 13.98 & 57.44   & 5.82 \\ 
   \hline
\end{tabular}
}
\end{table}

Haskell tends to perform the worst,
which is likely due to the use of STM,
which incurs an extra level of bookkeeping on every
operation.
Erlang performs better, but in general lags behind the other approaches.
C++/TBB tends to be the fastest, except in the
\bname{condition} and \bname{threadring} benchmarks,
which are both essentially single-threaded;
they are designed to test context switching overhead in various forms.
Go does quite well uniformly, never the fastest,
but never the slowest.
Lastly, {\qsname} performs mostly in line with C++/TBB and Go,
however it is the fastest in the condition benchmark.
In increasing order of geometric means:
C++/TBB (1.57s),
Go (1.82s),
{\qsname} (1.91s),
Erlang (5.01s), and
Haskell (12.20s).

\subsection{Summary}
This evaluation presents a wide variety of approaches
to concurrency and situates {\qsname} among them.
In particular,
we can see that {\qsname} is generally quite comparable
on coordination or concurrency tasks,
falling in the middle of the pack after Go and C++/TBB,
but faring better than Erlang and Haskell.
Note, however, that neither Go nor C++/TBB offers
any of the guarantees of {\qsname},
and {\qsname} offers more guarantees than Erlang.

For parallel problems, {\qsname} ranks 4th with all communication
burdens accounted for, but 1st when only computation is considered.
Again, we present both of these results only to say that
we expect real world usage to be between these extremes;
see the \bname{chain} benchmark for an indication of performance
on tasks which are not dominated by communication.
In any case, it still remains that {\qsname} outpaces Erlang in all cases,
and can provide reasonable performance comparable with the other approaches,
while still providing more guarantees in the execution model.

\begin{sloppypar}
For all problems, concurrent and parallel,
the geometric means are:
C++/TBB (0.71s),
Go (1.02s),
{\qsname} (1.61s),
Haskell (3.30s), and
Erlang (9.51s).
This places {\qsname} as the best performing of the
safe languages.
\end{sloppypar}
}
{
  \input{language_eval_short}
}

%}}}

%{{{ Related Work
\section{Related work}\label{sec:related-work}

Finding runtime and compiler optimizations is a vital research goal
when developing programming approaches for concurrency and
parallelism. While approaches in this area are based on a broad
variety of concepts, and in this respect each require different
solutions, this work profited from insights and discussions of a
number of related works.

Cilk~\cite{Blumofe:1995:CEM:209936.209958} is an approach to
multi-threaded parallel programming based on a runtime system that
provides load balancing using dynamic scheduling through work
stealing. Work stealing~\cite{blumofe:1994:scheduling} assumes the
scheduling forms a directed acyclic graph. In contrast, we tolerate
some cyclic schedules through the use of queues.
Since we use queues,
handler A can log work on handler B while handler B
logs work on A,
as long as they do not issue queries on one another (forcing a join edge).
We are not strict: edges go into handlers from the outside, other than at
spawn; this is actually the normal case when logging calls.
Although Cilk has been extended into Cilk++~\cite{frigo:2009:cilkxx},
this does not indicate a significant uptake of 
object-oriented concepts to ensure correctness properties such as race freedom.

X10~\cite{Charles:2005:XOA:1094811.1094852} is an object-oriented
language for high performance computing based on the partitioned
global address space model, which aims to combine distributed memory
programming techniques with the data referencing advantages in
shared-memory systems.
Although there is a mechanism to ensure local atomicity
through the keyword \texttt{atomic},
it is opt-in, and as such admits programs with data races by default.
The \texttt{async} blocks allow computations to run on different
address spaces, but there is no way for the caller to ensure
consistency between \texttt{async} blocks directed to the same address space.
The help-first stealing
discipline~\cite{guo:2009:work} in X10 offers that the spawned task is
left to be stolen, while the worker first executes the continuation;
this is in contrast to Cilk's work-first strategy where the spawned
task is executed first. The help-first strategy has benefits as it
avoids the necessity of the thieves synchronizing. This only applies
because the thefts in a finish block in X10 are serialized in work-first,
whereas they are not for help-first. This technique would not be
directly applicable to our work because {\qsname}
waits only on the result of a single handler.

Aida~\cite{Lublinerman:2011:delegated_isolation} is an execution model that,
like SCOOP, associates threads of control with portions of the heap.
The technique is implemented on top of Habanero-Java~\cite{Cave:2011:habanero_java},
an extension of the X10 implementation for Java.
When there is contention for a particular heap location,
the ``loser'' rolls back its heap modifications,
suspends, and appends itself (delegates) to the run queue of the winner,
effectively turning two concurrent tasks into a single one.
This is fundamentally different from the SCOOP model,
which also has isolated heaps,
but allows interaction between threads of control,
and even provides reasoning guarantees on this interaction.
Therefore the underlying mechanisms are fundamentally different,
where Aida requires efficient heap ownership and conflict resolution
via a parallel union-find algorithm,
SCOOP/Qs requires efficient communication which is attained via
novel and nested uses of specialized queue structures.
Otello~\cite{Zhao:2013:isolation_nested_parallelism} extends
the isolation found in Aida to include support for nested tasks.

\begin{sloppypar}
Another object-oriented approach which, like SCOOP,
associates threads of execution with
areas of the heap is JCoBox~\cite{schafer:2010:jcobox}.
It also makes the distinction (similar to \eif{separate})
between references that are local and those that are remote,
although this can only be applied per-class,
not per-object as in SCOOP.
Each CoBox contains a queue for incoming asynchronous calls,
though the reasoning guarantees are weaker for JCoBox,
so this structure can be simple.
The synchronous calls in JCoBox are also executed locally,
but no dynamic or static method to reduce
communication, ensuring data race freedom, is performed.
\end{sloppypar}

% In addition, OpenMP~\cite{openmp_board:2013:openmp_spec} is an approach to concurrency
% much in the style of TBB: providing high level constructs for common patterns.
% Essentially an extension to C/C++, rather than a library like TBB, 
% it provides pragmas to enable structured concurrent execution of sections of code,
% such as for array reductions and maps using for-loops.
% It does not however provide any reasoning guarantees and
% it is still the responsibility of the programmer to ensure that
% there are no data races.

Kilim~\cite{Srinivasan:2008:KIA:1428508.1428517} is a
framework that supports the implementation of Actor-based approaches
in Java. It improves message-passing performance by treating messages
differently from other Java objects, in that they are free of internal
aliases and owned by at most one Actor at a time.
The messages arrive via explicitly declared mailboxes in the objects,
which also do not provide the reasoning guarantees between messages
that the SCOOP model provides.
The Kilim mailboxes have, therefore, a more simplistic behaviour compared
to the queue-of-queues approach in {\qsname}.
Kilim also sets new
standards in creating lightweight threads, which are not tied to
kernel resources, thereby providing scalability and low context
switching costs. SCOOP implementations have previously been based on
operating system threads, and using lightweight threads in {\qsname} we
can report similar improvements in scalability as observed by Kilim.
Kilim is extended with ownership-based memory isolation~\cite{gruber:2013:ownership_isolation}
for messages to reduce the amount of unnecessary copying.
Although not strictly a message-based model,
{\qsname} may be able to apply this technique to so-called
expanded classes, which are more like standard C structures,
and are presently copied when used as arguments to separate calls.

We summarize the above approaches by stating  whether they offer
\emph{guards} (protection against races) and
\emph{delegation} (ability for one entity to give work to another).
\begin{itemize}
\setlength{\itemsep}{0.3ex}
\item \emph{No guarding, no delegation} -- Cilk/Cilk++.
\item \emph{Partial guarding, delegation} -- X10 allows delegation as
  the only way for one place to modify another.
  However, a place can asynchronously modify itself using 
  the same mechanism,
  thus there may be races within a place.
\item \emph{Guarding, protective delegation} --
  Aida and Otello extend X10 with the ability to resolve races by
  rolling back changes and reducing the amount of concurrent execution.
\item \emph{Guarding, delegation} --
  JCoBox and Kilim both have different approaches to the actor/active
  object model. This implies strict guarding and delegation of actions.
\item \emph{Guarding, enhanced delegation} --
  SCOOP follows the actor approach, but also offers enhanced
  delegation by allowing clients to maintain pre/postcondition reasoning with
  the handlers that they are delegating to.
\end{itemize}

%}}}

%{{{ Conclusion
\section{Conclusion}\label{sec:conclusion}

We have presented {\qsname}, an efficient execution model and
implementation for the SCOOP concurrency model. As many other
programming models that ensure strong safety guarantees, SCOOP
introduces restrictions on program executions, which can become
performance bottlenecks when implemented naively, standing in the way
of practicality and more widespread adoption. The key to our approach
was a reformulation of the SCOOP guarantees in abstract form, allowing
one to explore a larger design space for runtime and compiler
optimizations than previous operational descriptions. In particular,
it enabled us to remove much of the need for synchronization between
threads, thereby providing more opportunities for parallelism. In the
evaluation of our approach, we traced the impact of the key
optimizations, and compared {\qsname} with a number of well-known and
varied approaches to concurrency and parallelism: C++/TBB, Go,
Haskell, and Erlang. This confirmed that, on a broad benchmark
including both coordination-intensive and computation-intensive tasks,
{\qsname} can compete with and often outperform its competitors.

SCOOP offers a method of controlling access to other
actors which is more exclusive than typical Actor-like languages.
In SCOOP, messages can be bundled together to provide
better pre/postcondition reasoning between messages (calls).
The underlying techniques used in {\qsname} are an efficient
way to offer temporary control of one active object, or actor,
over another.
As such the technique could also be used in approaches like
JCoBox~\cite{schafer:2010:jcobox} or
Kilim~\cite{Srinivasan:2008:KIA:1428508.1428517} to
provide more structured interactions between
entities.

In the future, we plan to further explore the utility of the
private queue design, in particular the usage of sockets
as the underlying implementation.
To further investigate and advance the efficiency of the runtime,
a SCOOP-specific instrumentation for the runtime,
providing detailed measurements for the internal components,
will be essential.

\section{Acknowledgments}
This work was supported by ERC grant CME \#291389.

%}}}

\bibliographystyle{abbrv}
\bibliography{bibfile}

\end{document}